\documentclass[twocolumn,twocolappendix]{aastex631}
\usepackage{CJK}

\usepackage{amsmath}
\usepackage{appendix}
\usepackage{enumitem}

\DeclareGraphicsExtensions{.eps,.eps.gz,.epsi}
\newcommand{\teff}{\mbox{$T_{\rm eff}$}}
\newcommand{\logg}{{\rm{log}~$g$}}

\newcommand{\feh}{{\rm [Fe/H]}} 
\newcommand{\ebv}{$E(B-V)$}

\shortauthors{Huang et al.}

\graphicspath{{./}{my_figures/}}

\begin{document}
\begin{CJK*}{UTF8}{gbsn}
\title{Recalibration of the Landolt UBVRI Standard Stars and \\ 
the Generation of 5.4 Million New UBVRI Standard Stars using LAMOST and Gaia}
\author[0000-0002-1259-0517]{Bowen Huang (黄博闻)}
\affiliation{Institute for Frontiers in Astronomy and Astrophysics, Beijing Normal University, Beijing, 102206, China}
\affiliation{School of Physics and Astronomy, Beijing Normal University No.19, Xinjiekouwai St, Haidian District, Beijing, 100875, China}

\author[0000-0003-2471-2363]{Haibo Yuan (苑海波)}
\affiliation{Institute for Frontiers in Astronomy and Astrophysics, Beijing Normal University, Beijing, 102206, China}

\affiliation{School of Physics and Astronomy, Beijing Normal University No.19, Xinjiekouwai St, Haidian District, Beijing, 100875, China}

\correspondingauthor{Yuan Haibo (苑海波)}
\email{yuanhb@bnu.edu.cn}

\author[0000-0001-8424-1079]{Kai Xiao (肖凯)}
\affiliation{School of Astronomy and Space Science, University of Chinese Academy of Sciences, Beijing, 100049, China}

\author[0000-0003-1863-1268]{Ruoyi Zhang (张若羿)}
\affiliation{Department of Astronomy, Tsinghua University, Beijing 100084, China}
\affiliation{Institute for Frontiers in Astronomy and Astrophysics, Beijing Normal University, Beijing, 102206, China}
\affiliation{School of Physics and Astronomy, Beijing Normal University No.19, Xinjiekouwai St, Haidian District, Beijing, 100875, China}

\begin{abstract} 
We present an independent validation and recalibration of the Landolt 2013 (celestial equator and $\delta \sim -50^\circ$) and 2016 ($\delta \sim -50^\circ$) standard stars in the Johnson $UBV$ and Kron-Cousins $RI$ systems, using tens of thousands of XPSP data from the BEst STar (BEST) database.
Our analysis reveals an overall zero-point offset between the 2016 and 2013 datasets. We further identify zero-point offsets for each standard field, ranging from 5 -- 14 mmag across all $UBVRI$ bands, with correlations between offsets in different bands. Additionally, we confirm the spatial structures up to 7 -- 10 mmag in the $BVRI$ bands.
We also find that spatial structures are similar across bands for the same field, and similar across different fields for the same band. These similarities may arise from the averaged flat-fields from each observing run.
The recalibrated results are consistent with the XPSP data within 48 mmag in the $U$ band, 11 mmag in the $B$ band, and 5 -- 6 mmag in the $VRI$ bands in the brightness $G<16$. 
Furthermore, based on stellar atmospheric parameters from LAMOST DR12 and Gaia DR3 photometry, along with the XPSP data, we derive temperature- and extinction-dependent extinction coefficients for the $UBVRI$ bands as well as a LAMOST \& Gaia-based catalog of 5.4 million standard stars in the $UBVRI$ bands, for which the U-band photometry of the vast majority of sources exhibits significantly higher precision than XPSP.
The recalibrated Landolt standard stars and LAMOST \& Gaia-based standard stars will be available on the BEST website (\url{https://nadc.china-vo.org/data/best/}) and (\url{https://doi.org/10.12149/101704}).

\end{abstract}


\section{Introduction} 
Astronomy is observation-driven, and our understanding of the physical properties of astronomical objects depends on the precision of our measurements. Nowadays, with the rapid growth of wide-field photometric survey projects, we are provided with vast amounts of photometric data. To accurately measure the physical properties of astronomical objects, it is essential to ensure the consistency of flux measurements across widely separated targets, varying observational conditions, different detector positions, and multiple observing times. Therefore, high-precision photometric calibration is critical for conducting reliable scientific research.

Traditionally, the calibration of photometric measurements has been based on standard stars. As a widely used broad-band photometric system, the Johnson-Kron-Cousins (JKC) UBVRI standard sequences have been established through a series of studies (\citealt{Johnson1953UBVdefinition,Kron1953RI,Johnson1954UBVrefine,Johnson1955UBVrefine,Johnson1966RI,Cousins1976RImodified,Landolt1973,Landolt1983,Landolt1992a,Landolt2007b,Landolt2009,Landolt2013}). 
Over the past few decades, numerous photometric investigations have relied on these so-called ``Landolt standard stars'' (e.g. \citealt{LandoltCite1:HSTWFPC2,LandoltCite2:SupernovaLegacySurvey,LandoltCite3:GALEX,LandoltCite4:SwiftUVOT}). However, the traditional standard star method no longer meets the demands of increasing photometric calibration precision. Due to the limited number and precision of available standard stars, achieving a calibration precision better than 1 per cent for ground-based observations remains a significant challenge (e.g. \citealt{StubbsTonry1percent}). 

In recent years, a series of photometric calibration methods have been proposed, including both ``hardware/observation-driven'' and ``software/physics driven'' methods (\citealt{Huang2022bReview}), gradually breaking through the 1 per cent barrier.
Among these methods, the stellar color regression (SCR) method and Gaia BP/RP (XP) spectra-based synthetic photometry (XPSP) have yielded remarkable results for photometric calibration by generating a substantial number of high-precision standard stars. 

The SCR method, first proposed by \cite{S82Yuan2015a}, demonstrates that intrinsic colors can be precisely determined from a few physical quantities, such as stellar atmospheric parameters, normalized stellar spectra, colors, and the combination of color and metallicity (\citealt{YL2021,Huang2022aS82,Xiao2023b,YL2024}).
With the release of extensive stellar atmospheric parameters from large-scale spectroscopic surveys such as the Large Sky Area Multi-Object Fiber Spectroscopic Telescope (LAMOST; \citealt{LAMOSTCui2012,LAMOSTDeng2012,LAMOSTZhang2012,LAMOSTLiu2014}) along with Gaia Data Release 3 (DR3) photometry (\citealt{Gaia2023bDR3content}), millions of standard stars are available.   
Many prominent photometric surveys have been calibrated using the SCR method, robustly break through the 1 per cent barrier. These include the Sloan Digital Sky Survey (SDSS; \citealt{SDSS2000}) Stripe 82 (\citealt{S822007}; calibration: \citealt{S82Yuan2015a} and \citealt{Huang2022aS82}); Gaia Data Release 2 (\citealt{GaiaDR22018}) and Early Data Release 3 (EDR3; \citealt{GaiaEDR32021a}; calibration: \citealt{Niu2021a,Niu2021b} and \citealt{YL2021}); the SkyMapper Southern Survey (SMSS; \citealt{SkyMapper2018DR1}) Data Release 2 (\citealt{SkyMapper2019DR2}; calibration: \citealt{SkyMapper2021DR2HY}); Pan-STARRS1 (PS1; \citealt{PS12012}; calibration: \citealt{Xiao2022PS1} and \citealt{Xiao2023b}); and the Stellar Abundance and Galactic Evolution Survey (SAGES; \citealt{SAGES2018,SAGES2019,SAGES2023DR1,SAGES2025gri}; calibration for gri bands: \citealt{Xiao2023a}).

In June 2022, absolutely calibrated XP spectra for approximately 220 million sources were released as part of Gaia DR3 (\citealt{XPInternalCali}, \citealt{Gaia2023bDR3content}, \citealt{XPInteralCaliDe}, \citealt{XPExternalCali}), covering wavelengths from 336 to 1020\,nm and mostly for sources with magnitudes of $G < 17.65$. This dataset provides an unprecedented opportunity to synthesize photometry in any desired band within this wavelength range (XPSP; \citealt{Gaia2023aXPSP}).
However, the XP spectra suffer from systematic errors associated with magnitude, color, and extinction, especially over the wavelength range of 336 -- 400\,nm (\citealt{XPExternalCali}; \citealt{HBW2024a}). To address this, \cite{HBW2024a} performed comprehensive corrections to the Gaia XP spectra. This correction was subsequently used to improve the XPSP method (\citealt{Xiao2023c}), which no longer relies on the 343 coefficients and is capable of directly deriving multi-band magnitudes with higher accuracy.

More recently, Xiao et al. (in prep.) created the BEst STar (BEST) database, comprising over 200 million all-sky standard stars along with well-calibrated composite catalogs. These high-precision photometric standard stars are established using the SCR and improved XPSP methods (\citealt{Xiao2023c}). However, due to the brightness limitations of the LAMOST and Gaia DR3 XP spectra, the standard stars are primarily concentrated at the bright magnitude range ($G < 18$), resulting in a shortage of standard stars for fainter sources.
Fortunately, Clem et al. conducted observations of a series of faint UBVRI standard star fields in the JKC photometric system in 2013 and 2016 (\citealt{Clem2013,Clem2016}; hereafter L13 and L16), covering a magnitude range in $V$ from 12 to 22. These fields provide valuable faint standard stars, and could be a supplement to the future BEST standard star set by extending the brightness coverage to fainter sources. 
Nevertheless, their photometric precision is limited by the constraints of ground-based calibration, with relatively large systematic errors in the $U$ band and on the order of a few per cent in the $BVRI$ bands.

In this study, we perform a photometric re-calibration of the Landolt Standard Star (L13 and L16) aiming to achieve uniform photometry with a precision at the milli-magnitude level using the BEST database. In addition, we assess the quality of the standard stars from \cite{Landolt1992a} (hereafter L92).
Furthermore, we utilize common sources between XPSP and LAMOST to derive temperature- and extinction-dependent extinction coefficients in the UBVRI bands. 
By applying these coefficients and incorporating the SCR method, we construct a catalog of 5.4 million standard stars in the UBVRI system, based on LAMOST stellar atmospheric parameters and Gaia DR3 Blue Photometer (BP) and Red Photometer (RP) photometry. This catalog will be integrated into the BEST database.

The structure of this paper is as follows:
Section 2 describes the dataset used in this work.
In Section 3, we provide the calibration of the L13 and L16 standard stars, followed by quality assessment for L92 standard stars.
In Section 4, we derive temperature- and extinction-dependent extinction coefficients in the UBVRI bands, and construct a catalog of 5.4 million standard stars with UBVRI-band photometry derived using the SCR method based on LAMOST stellar parameters and Gaia BP/RP photometry.
A summary is presented in Section 5.

\section{DATASETS} 
\subsection{Landolt Stnadard Stars} 
\cite{Landolt1992a} presented a catalog of 526 UBVRI photometric standard stars in the magnitude range $11.5 \lesssim V \lesssim 16$ and color range $-0.3 \lesssim B-V \lesssim 2.3$, located near the celestial equator. All data were obtained using photomultiplier observations. Among these, 81 stars were previously included as standard stars in \cite{Landolt1983}, and their final magnitudes and colors were determined by taking a weighted average of the values from both \cite{Landolt1983} and \cite{Landolt1992a}.

Subsequently, \cite{Clem2013} published approximately 45,000 faint, high-quality CCD-based standard stars in the UBVRI system, distributed across 60 fields centered near the celestial equator and at $\delta \approx -50^{\circ}$. \cite{Clem2016} added approximately 2,000 additional stars at $\delta \approx +50^{\circ}$. Together, these 47,000 stars are referred to hereafter as the Landolt stars. Most of them fall within the magnitude range $12 \lesssim V \lesssim 22$ and the color range $-0.3 \lesssim B-V \lesssim 1.8$, with each star having, on average, 67 measurements in each of the $UBVRI$ filters. The typical photometric uncertainties are approximately 15 mmag in the U band,  7.5 mmag in the B band and 3 -- 4 mmag in the VRI bands.

\subsection{Gaia Data Release 3}
The ESA Gaia mission is primarily dedicated to obtaining astrometry, photometry, and spectroscopy for objects within the Milky Way and the Local Group (\citealt{GaiaMission2016a}).
Gaia DR3 includes over 1.8 billion sources with G and BP/RP broadband photometry, and 1.47 billion sources with at least five-parameter astrometric solutions (i.e., two positions, parallax, and two proper motion components; \citealt{Gaia2023bDR3content}). It also provides approximately 220 million XP spectra covering the wavelength range 336–1020\,nm.
Instead of providing flux as a function of wavelength, Gaia XP spectra are represented as projections onto 55 orthonormal Hermite basis functions for both the BP and RP channels (\citealt{XPInternalCali}; \citealt{XPInteralCaliDe}; \citealt{XPExternalCali}).

However, the XP spectra have been found to suffer from systematic errors that depend on the spectral energy (described by brightness, color and extinction; \citealt{XPExternalCali}; \citealt{HBW2024a}). \cite{HBW2024a} presented comprehensive corrections for these systematic errors in the wavelength domain, valid over the ranges $-0.5 < BP-RP < 2$, $3 < G < 17.5$, and $\ebv < 0.8$. Validation tests demonstrated that these corrections significantly reduce the systematic errors, achieving an internal precision of 1 -- 2 per cent.

\subsection{LAMOST Data Release 12} 
The Large Sky Area Multi-Object Fiber Spectroscopic Telescope (LAMOST), operated by the National Astronomical Observatories of China, Chinese Academy of Sciences, is a 4-meter quasi-meridian reflective Schmidt telescope equipped with 4000 fibers. The LAMOST Data Release 12 (DR12)\footnote{\url{https://www.lamost.org/dr12/}} includes a total of 12,602,390 low-resolution spectra collected over the past ten years, covering the full optical wavelength range of 369 -- 910\,nm with a spectral resolving power of $R \sim 1800$.
Based on these low-resolution spectra, DR12 also provides atmospheric parameters for approximately 8,370,041 AFGK-type stars, including effective temperature ($\teff$), surface gravity (\logg), and metallicity ($\feh$).
These fundamental parameters are derived using the LAMOST Stellar Parameter Pipeline \citep{LASP, LAMOSTLuo2015}. For spectra with signal-to-noise ratios greater than 20, the typical internal uncertainties are approximately $\teff \sim 50-100$\,K, \logg\ $\sim 0.05-0.1$\,dex, and $\feh \sim 0.05-0.1$\,dex \citep{Niu2021a}. The corresponding sources span a broad range in color and magnitude, with $0 < BP-RP < 3$ and $9 < G < 17.8$.

\subsection{BEST Standard stars in the JKC System} 
The BEST database (K. Xiao et al. 2025, submitted) provides over 200 million XPSP standard stars, derived using an improved XPSP method (\citealt{Xiao2023c}) based on the corrected Gaia XP spectra (\citealt{HBW2024a}). 
The adopted transmission curves of the BEST database in the JKC system are taken from \cite{Bessell2012}, and the corresponding magnitudes are obtained by convolving the fluxes of the corrected Gaia XP spectra with these transmission curves. For further details, see Section 3.2 of \cite{Xiao2023c}.
These standard stars are distributed across the entire sky in the AB-magnitude system (\citealt{Oke1983}).
Overall offsets (arising from differences between the AB magnitude system and the VEGAMAG system (\citealt{Johnson1953UBVdefinition})), minor brightness-dependent differences, and intermediate color-dependent differences between the BEST standard stars and the Landolt standard stars are standardized through the L13 dataset and an iterative process of the calibration, as described in the following sections. The final standardized BEST standard stars in the VEGAMAG system will also be released as part of the BEST database, in consideration of the widespread adoption of Landolt standard stars.

\section{Calibration of Landolt Standard Stars} 
The calibration strategy consists of three components for each band, carried out in three iterations:
1) standardization of the BEST standard stars based on the L13 dataset;  
2) determination of the zero-point offsets for each field;  
3) characterization of spatial structures for each field.
The final calibrated magnitudes of the Landolt standard stars are obtained using the following equation:
\begin{equation}
    m_{x}^{cali} = m_{x}^{L13/L16} + \delta m _{x}^{ZP} + \delta m _{x}^{SS}
    \label{EQ1}
\end{equation}
while $x$ denotes the UBVRI bands. Here, $m_{x}^{cali}$ is the calibrated magnitude, $m_{x}^{L13/L16}$ is the original magnitude from L13 or L16, $\delta m _{x}^{ZP}$ represents the zero-point correction for each field, and $\delta m _{x}^{SS}$ accounts for the spatial structure correction within each field. It is worth noting that $\delta m _{x}^{ZP}$ is a single value for each field/filter combination. In contrast, $\delta m _{x}^{SS}$ is defined for each individual star, derived from the standard stars within a fixed radial range around it. The radius is fixed for each field, and the same set of standard stars may be repeatedly used for neighboring stars. The adopted radius typically ranges from 0.04 to 0.175 degrees, depending on the number of available standard stars.

\subsection{Data Quality Selection}
We combine the L13 and L16 photometric data with the BEST standard stars by cross-matching within a radius of 1 arcsecond. This yields 19,875 and 702 common sources for L13 and L16, respectively. For the subsequent quality assessment of the L92 standard stars, a larger matching radius will be adopted due to the significant epoch difference.
To ensure the reliability of both data quality and standardization in the calibration process, the following criteria are applied to all common sources:

\begin{enumerate}[itemsep=0.2cm, topsep=0.5cm, parsep=0.2cm, partopsep=0.3cm]
    \item For L13 and L16, the uncertainties in the $U$, $B$, $V$, $R$, and $I$ magnitudes are required to be smaller than 0.01, 0.01, 0.0075, 0.0075, and 0.0075 mag, respectively.
    \item For the BEST standard stars, the uncertainties in the $U$, $B$, $V$, $R$, and $I$ magnitudes are smaller than 0.3, 0.02, 0.005, 0.005, and 0.005 mag, respectively.
    \item Photometric and color constraints: $11 < G \le 17.65$, $0 \le BP-RP \le 1.5$ and $B-V \le 1.5$
    \item $phot\_bp\_rp\_excess\_factor <0.023 \times (BP-RP)^2+0.055 \times (BP-RP)+1.165$. The $phot\_bp\_rp\_excess\_factor$ is defined as $C = \frac{I_{BP}+I_{RP}}{I_{G}}$ to evaluate the background and
contamination issues in both $BP$ and $RP$ photometry and spectra. Our empirical cut helps to exclude sources with low-quality Gaia measurements. Furthermore, this criterion effectively removes any potential extended sources, such as galaxies.
\end{enumerate}

After applying criteria 1 -- 4, 11,822 standard stars remain for L13 and 297 for L16. 
In the L13 data set, 35 fields contain fewer than 100 standard stars, 19 fields contain 100 -- 500, and 7 fields contain more than 500; in particular, field LSE259 includes about 2,000 standard stars.
For L16, due to the limited number of standard stars, only fields containing more than ten standard stars are used to calculate and apply $\delta m _{x}^{ZP}$, while the remaining fields are not calibrated.

It should be noted that the above criteria are designed to select stars as reliably as possible for calibration. However, it is inevitable that L13 and L16 potentially contain a small number of variable stars, galaxies, some cases of poor photometry. Unfortunately, neither L13 nor L16 provide any quality indices, and we can only use the $phot\_bp\_rp\_excess\_factor$ to remove extended sources such as galaxies. Nevertheless, a small number of variable stars and instances of poor photometry do not affect the overall photometric calibration; these potential issues will be inherited in the final calibrated catalog.

\subsection{Standardization}
Due to potential differences between the transmission curves adopted by BEST (\citealt{Bessell2012}) and the actual transmission curves in L13, L16, and L92, which account for atmospheric absorption, the wavelength-dependent reflectivity of telescope mirrors, as well as the linear extrapolation of Gaia XP spectra at the blue end and slight residual systematic errors in the Gaia XP spectra, we perform a standardization of the BEST standard stars to correct for the average systematic deviations in each band introduced by these effects, thereby ensuring statistical consistency between the BEST standard stars and Landolt standard stars within a specific color-magnitude range.
It is noted that the standardization of the BEST standard stars is anchored to the largest dataset, L13, ensuring consistency between the standardized BEST stars and L13. Furthermore, the L16 dataset is subsequently calibrated to maintain consistency with L13 in the later calibration process.

\begin{figure*}[ht]
\includegraphics[width=180mm]{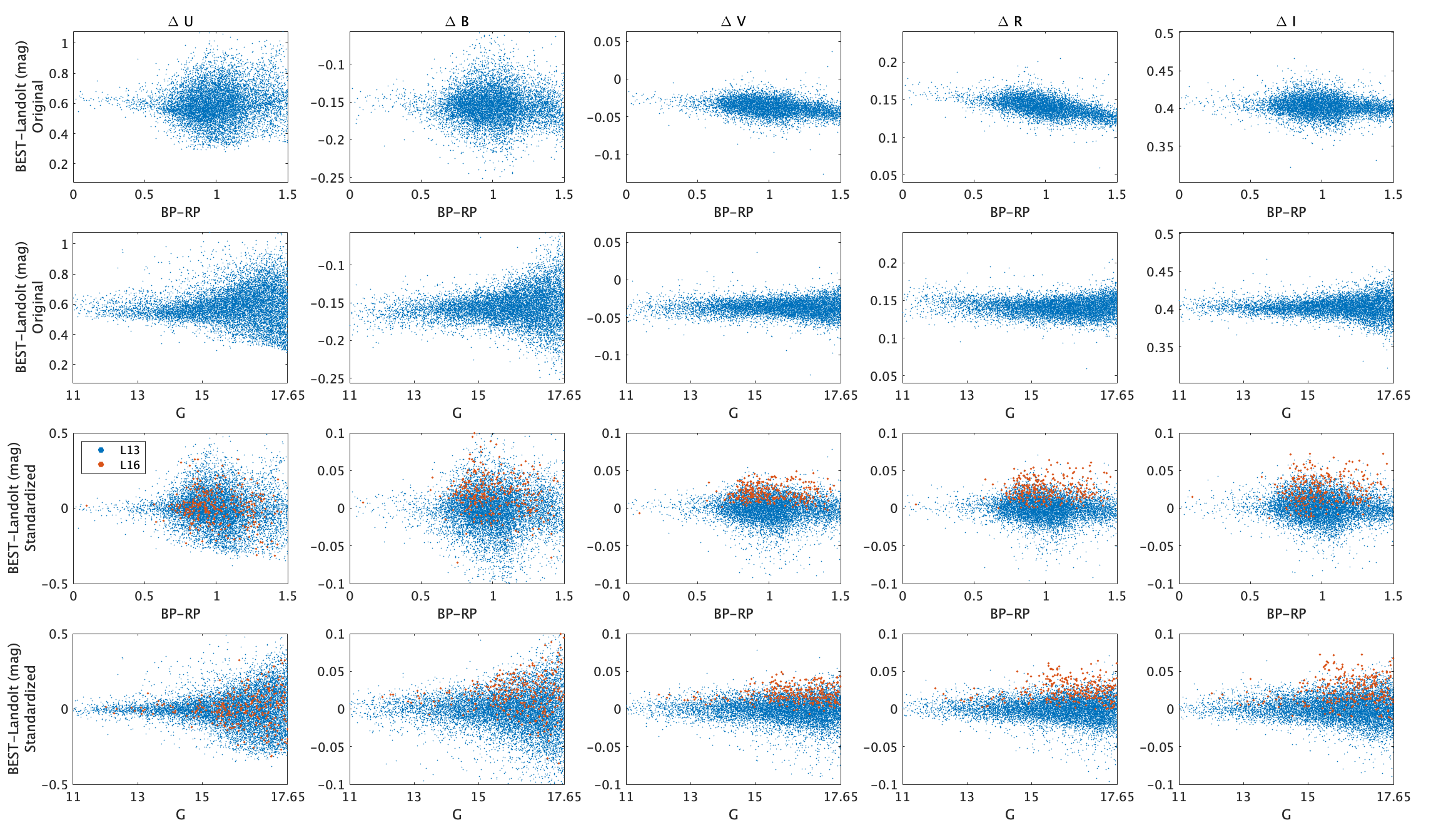}
\caption{The differences between the BEST standard stars and the L13/L16 datasets in the third calibration iteration, shown as a function of $BP-RP$ color and $G$ magnitude. The top two rows correspond to the original BEST standard stars, while the bottom two rows show the standardized BEST standard stars.} 
\label{CMcor}
\end{figure*}
We perform a bivariate polynomial fit to the differences between the original BEST standard stars and L13, using $BP-RP$ color and $G$ magnitude as independent variables, within the ranges $11 < G \leq 17.65$, $0 \leq BP-RP \leq 1.5$, and $B-V < 1.5$. To ensure the robustness and accuracy of the fit, this process is included in the calibration iterations, and some unreliable sources in the $U$ band are excluded by iterative 3-$\sigma$ clipping in the fit during the calibration iterations.

Figure~\ref{CMcor} displays the differences between the BEST standard stars (both original and standardized) and the L13/L16 datasets after the second calibration iteration, shown as a function of $BP-RP$ color and $G$ magnitude.
These results demonstrate that the color- and magnitude-dependent trends are effectively removed after the standardization process.

In addition, three noteworthy points should be mentioned. 
First, in the five UBVRI bands, the original BEST standard stars and L13 show significant overall offsets, with the offset values differing from band to band. These offsets arise from differences between the AB magnitude system and the VEGAMAG system (see the Table 3 of \citealt{Bessell2012}). 
Second, a noticeable overall offset also exists between L13 and L16, which will be corrected in the subsequent calibration by adjusting the zero-points of each field to ensure consistency between L13 and L16.
Third, due to the lack of detailed information on the observatories and instruments for each star in L13/L16, and given that variations in the transmission curves inherently introduce systematics dependent on stellar SEDs-which cannot be perfectly represented by a single color-the standardization can only statistically correct the overall systematic offsets within the specified color–magnitude range.

\subsection{Zero-Point Corrections for Each Field}
\begin{figure*}[htbp]
\includegraphics[width=180mm]{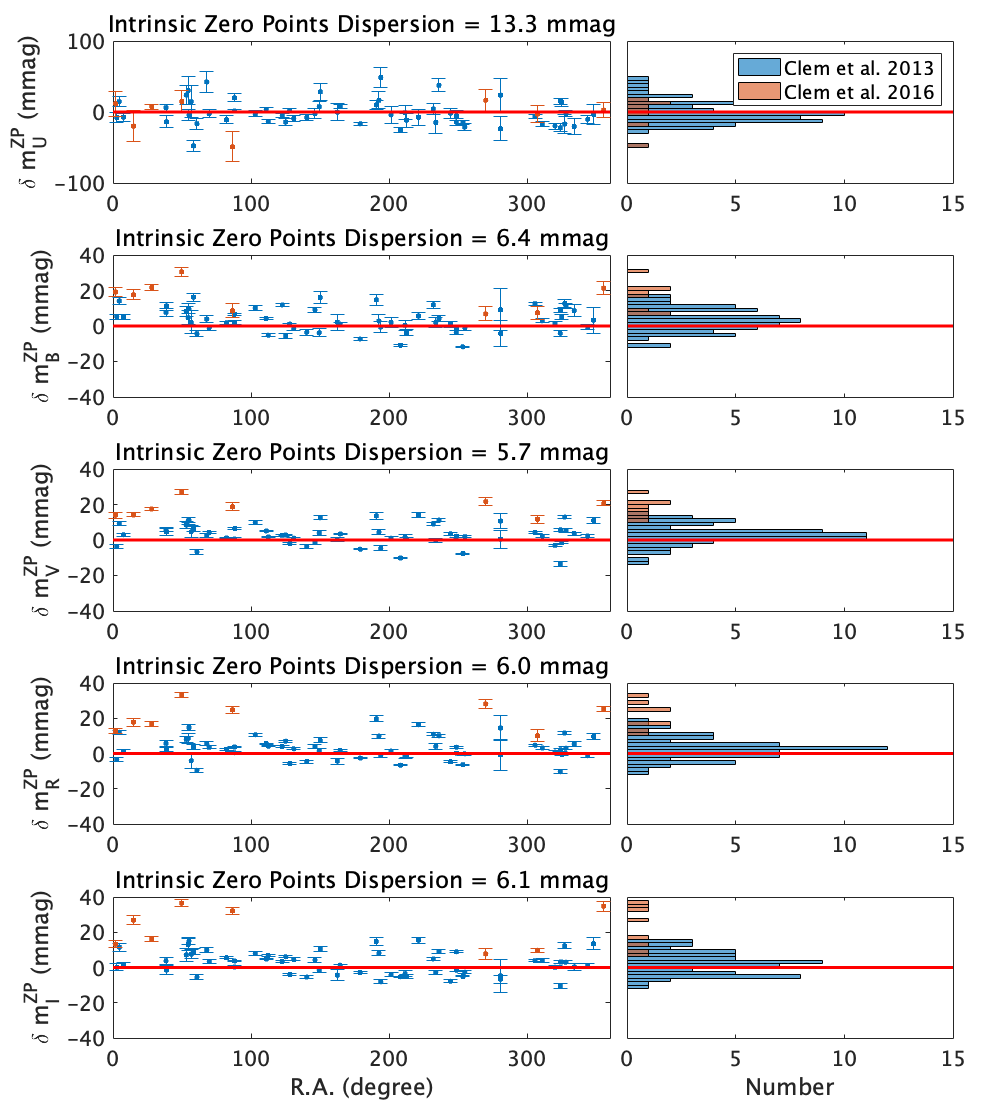}
\caption{The $\delta m_{x}^{ZP}$ values for each field are plotted against Right Ascension (R.A.). The panels from top to bottom correspond to the UBVRI bands. Blue points with error bars represent fields from L13, while red points with error bars represent those from L16. The panels on the right show the distribution of $\delta m_{x}^{ZP}$, along with the maximum likelihood estimation of the intrinsic dispersion for the UBVRI bands, based on the distribution and uncertainties of $\delta m_{x}^{ZP}$ in L13. } 
\label{FFZP_RA}
\end{figure*}

The zero-point corrections, $\delta m_{x}^{ZP}$, for each field are determined as the median of the magnitude differences between the original magnitudes of sources from L13 or L16 and the corresponding standardized BEST standard stars. The uncertainties in $\delta m_{x}^{ZP}$ are estimated based on the standard deviation of these differences and the number of sources used in the median calculation.
To infer the intrinsic dispersion of $\delta m_{x}^{ZP}$ (the original zero-point uncertainty of L13), we adopt a maximum likelihood approach that explicitly accounts for the observed dispersion in $\delta m_{x}^{ZP}$ and the associated measurement uncertainties. We assume that both the measurement uncertainties and the intrinsic scatter in $\delta m_{x}^{ZP}$ are independently drawn from Gaussian distributions. Figure~\ref{FFZP_RA} presents the values of $\delta m_{x}^{ZP}$ for each field as a function of Right Ascension (R.A.), along with their distributions. The panels from top to bottom correspond to the UBVRI bands, showing the intrinsic dispersion of $\delta m_{x}^{ZP}$ in each band. The intrinsic dispersion is approximately 13 mmag in the $U$ band, while it remains around 6 mmag for the other bands.  
And the maximum zero-point offsets among fields may reach approximately 49 mmag in the U band and around 15-20 mmag in the other bands.
It should be noted that all estimates of $\delta m_{x}^{ZP}$ are subject to uncertainties, and the maximum-likelihood approach based on the assumption of Gaussian distributions may deviate from reality. Therefore, the quantitative results inevitably involve some unavoidable uncertainties. As a reference, the standard deviations of current $\delta m_{x}^{ZP}$ distribution are higher by about 1.5 mmag in the U band and by less than 0.5 mmag in the BVRI bands compared to the current estimates.

\begin{figure*}[htbp]
\includegraphics[width=180mm]{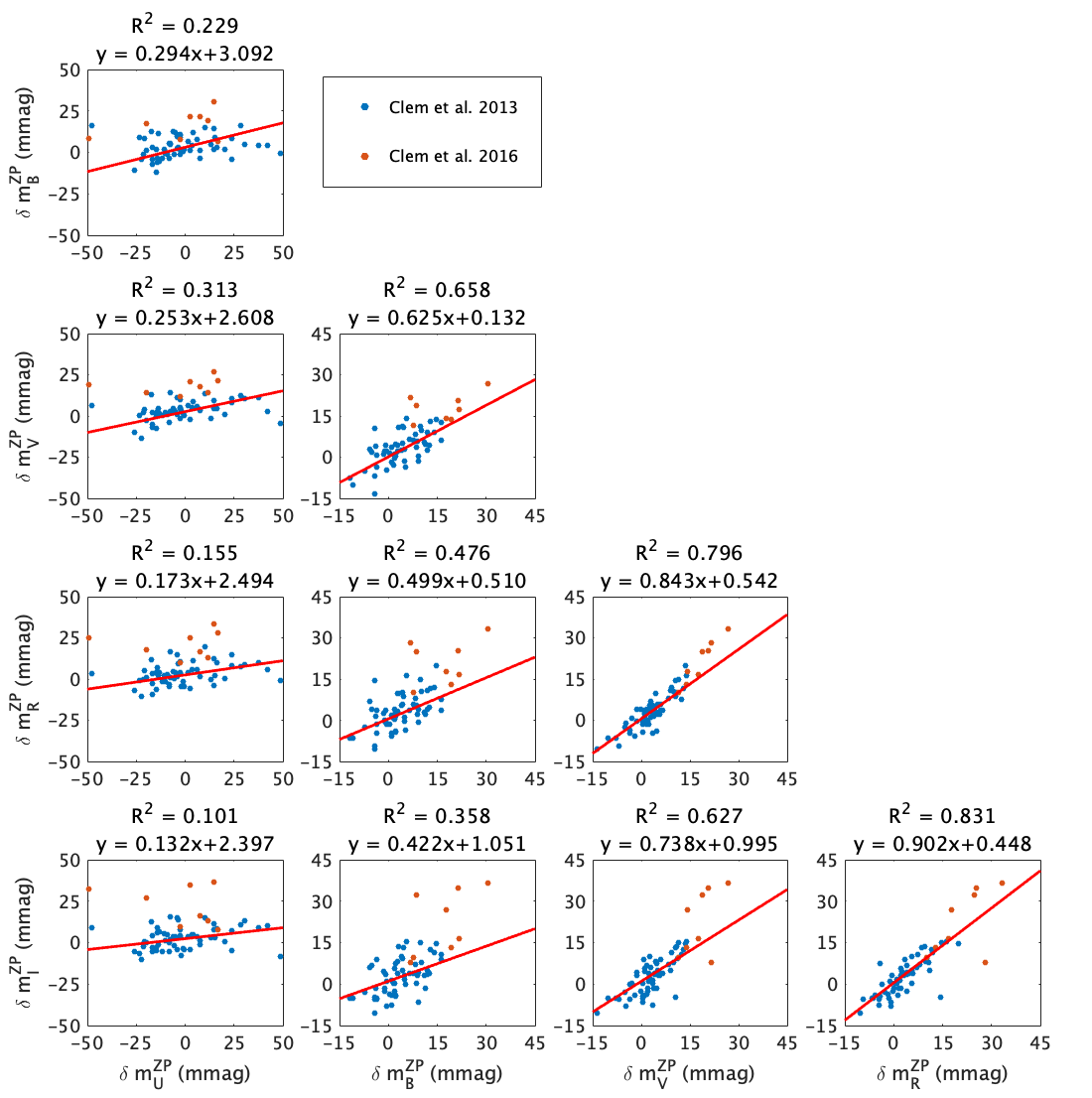}
\caption{The pairwise comparisons of $\delta m_{x}^{ZP}$ across the UBVRI bands. The solid red lines represent linear regression results based solely on fields from L13 and accounting for measurement uncertainties. The linear regression coefficients and coefficients of determination ($R^2$) are indicated in each panel.} 
\label{FFZP_rela}
\end{figure*}

Furthermore, based on the left panels, a correlation of $\delta m_{x}^{ZP}$ across different bands within the same field can be observed. We therefore further investigate and analyze this correlation.
The pairwise comparisons of $\delta m_{x}^{ZP}$ between the UBVRI bands are shown in Figure~\ref{FFZP_rela}. Based on the coefficients of determination ($R^2$) from the linear regressions, significant correlations are found between most band pairs, except those involving the $U$ band. In addition, stronger correlations are observed between bands with closer effective wavelengths.

The exact origin of these correlations remains uncertain. However, as noted in Section 3.4 of \cite{Clem2013}, the photometric calibrations were based on standard stars from \cite{Landolt2009}, observed with photomultipliers. Two factors may contribute to the observed correlations.
First, \cite{Landolt2009} updated and expanded the L92 (\citealt{Landolt1992a}) standard star catalog. As discussed in following (see Section 3.6), the quality assessment for L92 standard stars revealed similar correlations even for individual stars. The zero-point correlations in L13/L16 may therefore be inherited from these pre-existing correlations during the calibration process.
Second, in determining the zero-point for each band, air-mass dependent photometric transformations were applied (see the equations in Section 3.4 of \cite{Clem2013}). For some fields, temporal drifts in the photometric zero-point of up to $\sim\pm 0.02 ,\mathrm{mag},\mathrm{hr}^{-1}$ were reported. These correlations could also be combined result of temporal zero-point drifts and the photometric transformation process. The stronger correlations between bands with similar effective wavelengths are plausibly due to the similarity of their respective transformation relations.

\subsection{Spatial Structures Correction for Each Field}
\begin{figure*}[htbp]
\includegraphics[width=180mm]{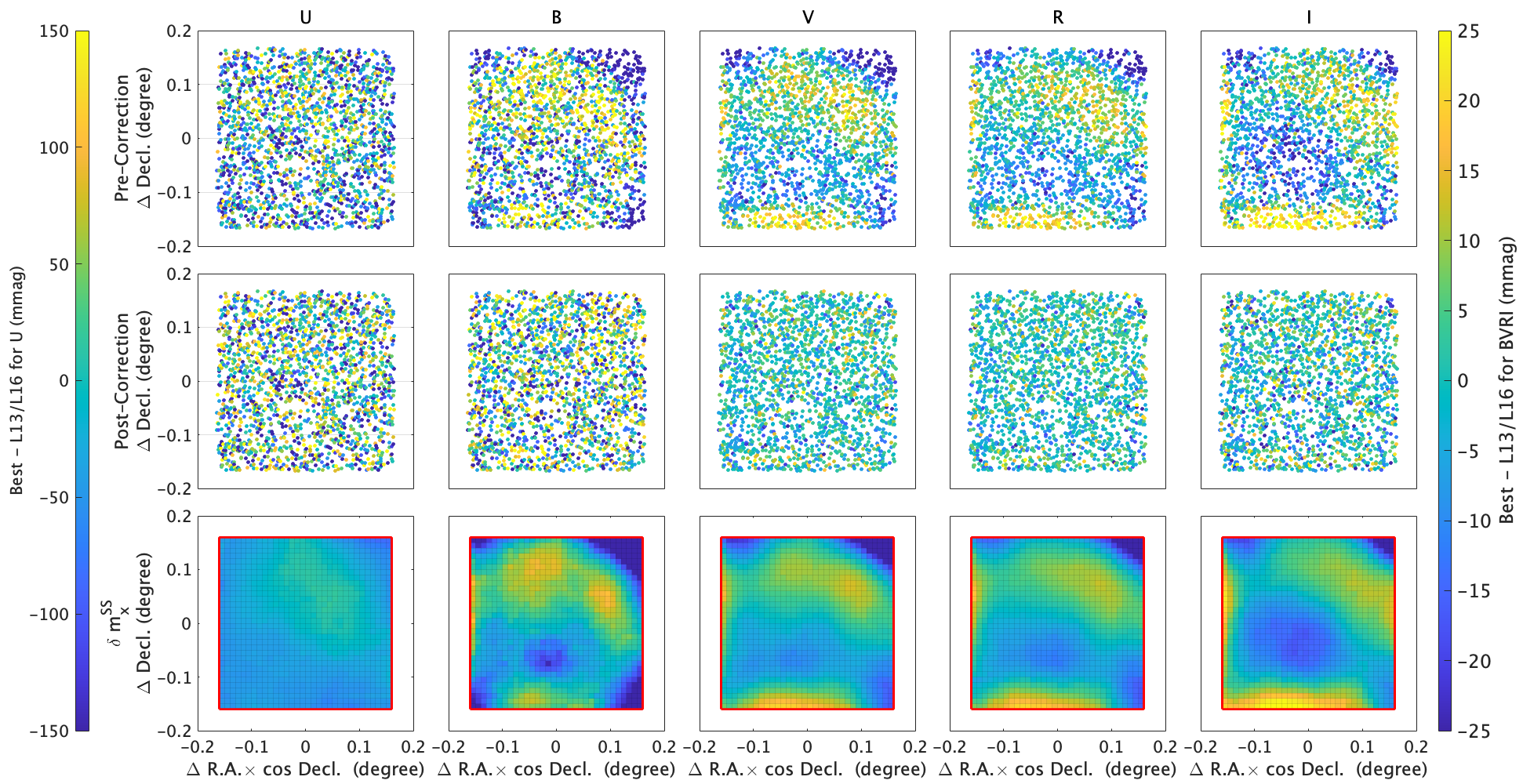}
\caption{A sample spatial structure correction for field LSE259. The first row illustrates the spatial distribution (relative to the field center) of magnitude differences between BEST standard stars and L13/L16 in UBVRI bands before spatial structure correction; the second row depicts these distributions after correction; the third row displays the characterized flat-field structure $\delta m_{x}^{SS}$, with red boxes marking the edges of the Y4KCam CCD.} 
\label{FF_sample}
\end{figure*}

After applying the zero-point corrections for each field as described earlier, we further derive the spatial structure correction for each field using the magnitude differences between the L13/L16 observations and the standard stars. Figure~\ref{FF_sample} shows an example based on the field with the largest number of standard stars (LSE259, observed with Y4KCam; \citealt{Clem2013}). A consistent ring-like pattern and signs of vignetting appear across all UBVRI bands, with particularly similar structures among the BVRI bands. 
We characterize and correct this structure using the numerical stellar flat-field method (discussed in \citealt{Xiao2024b}). The second and third rows of Figure~\ref{FF_sample} show the residuals after spatial structure correction and the reconstructed pattern of spatial structure $\delta m_{x}^{SS}$, respectively. The typical amplitude of the structures reach approximately 20 -- 30 mmag in the $U$ band and about 7 -- 10 mmag in the BVRI bands.

Not all fields contain a sufficient number of standard stars that are uniformly distributed across the field of view, which is essential for reliably characterizing spatial structures.
In fields with fewer standard stars, the numerical stellar flat-field method increases the pairing radius to incorporate a larger sample and preserve statistical robustness; however, this tends to capture only smoother spatial structures. Moreover, spatial structure correction is omitted for fields with very few standard stars or with markedly non-uniform spatial distributions.
Additionally, due to significantly larger uncertainties in both the $U$ band of the BEST standard stars and the L13/L16 photometry, spatial structure correction in the $U$ band is performed only for two fields -- LSE259 and WD0830-535, where standard stars are sufficiently abundant. In other fields, the $U$-band spatial structure cannot be reliably characterized and is therefore left uncorrected.

\begin{figure*}[htbp]
\includegraphics[width=180mm]{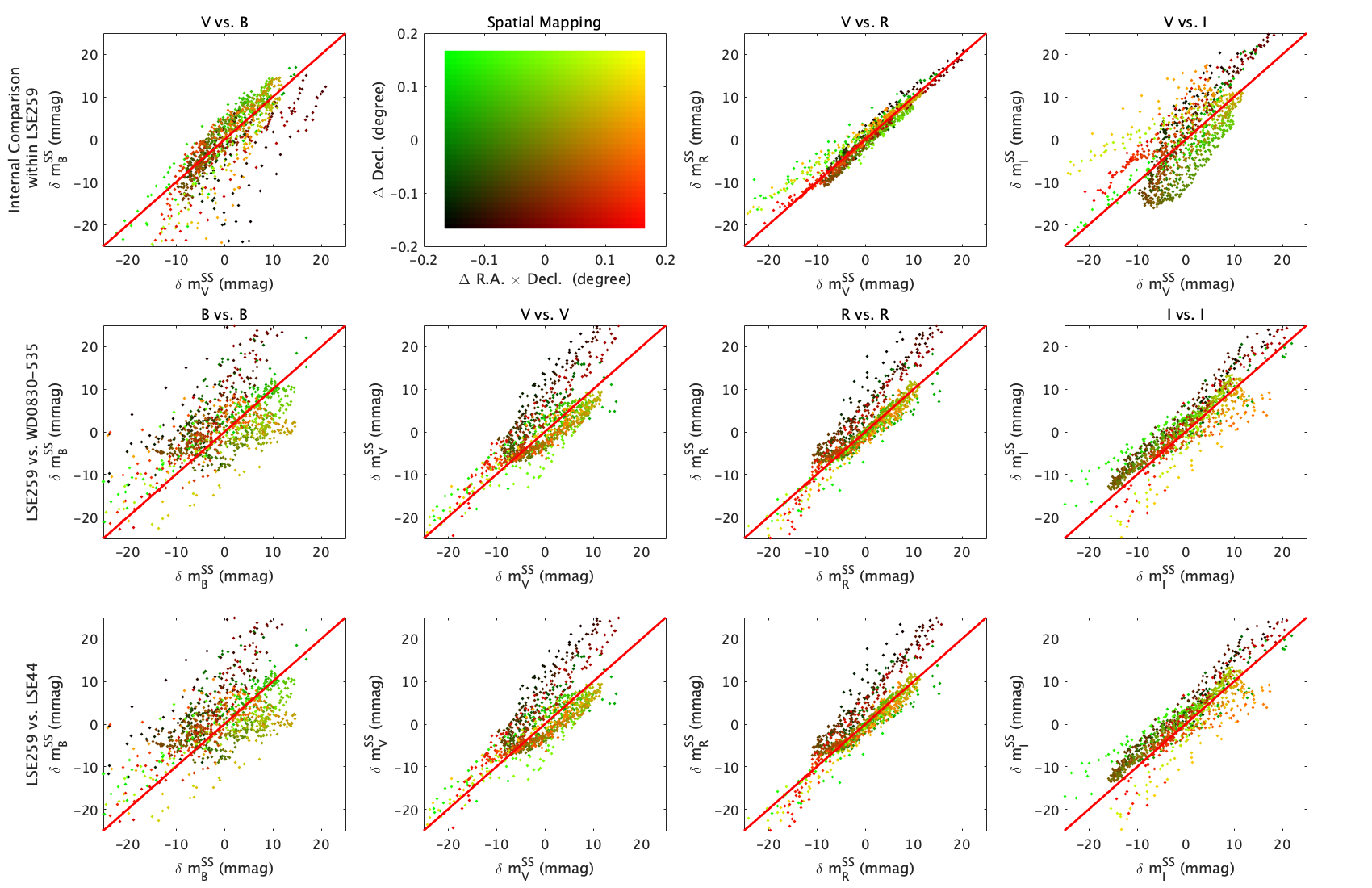}
\caption{Comparison of spatial structure pattern across different bands within LSE259 and between LSE259 and other fields.
The first row presents an internal comparison of spatial structure pattern among different bands within LSE259. The second and third rows show inter-field comparisons of corresponding bands between LSE259 and WD0830-535, and between LSE259 and LSE44, respectively. The solid red line marks the line of equality. The color of each point represents its position relative to the field center, as defined by the spatial mapping shown the panel at top row, second column.} 
\label{FF_compare}
\end{figure*}

Notably, when a sufficient number of standard stars are available, similar ring-like structures and vignetting patterns in the spatial structure corrections from the Y4KCam detector can be consistently identified across different fields.
To further verify this, we compare the patterns of spatial structures in the BVRI bands for the three fields with the largest number of standard stars. These comparisons include an internal comparison among different bands within LSE259, as well as inter-field comparisons of corresponding bands between LSE259 and WD0830-535. As shown in Figure~\ref{FF_compare}, the patterns of spatial structures among different bands demonstrate substantial structural consistency, which also extends across different fields. Although the B band comparisons show slightly larger scatter, and the correlation varies with position on the CCD, the spatial structure corrections $\delta m_{x}^{SS}$ exhibit clear spatial correlation-particularly for points located at similar CCD positions (i.e., with similar colors).
These results indicate that, at least for data obtained with Y4KCam, the flat-field patterns remain consistent across both bands and fields, with a typical amplitude of approximately 7 -- 10 mmag.

This consistency is understandable. According to the Y4KCam observing logs\footnote{\url{https://www.astronomy.ohio-state.edu/Y4KCam/}} for L13, twilight flats for different filters are taken consecutively. Although the exact cause remains uncertain, the sequential observation of twilight flats across filters likely accounts for the observed inter-band similarity in flat-field patterns.
Furthermore, as described in Section 2.3 of \cite{Clem2013}, twilight flats taken with Y4KCam typically yield the best results within $\sim$ 1 -- 2 per cent, which is consistent with the typical amplitudes found in our results.
However, it is difficult to collect a sufficient number of high-signal flats within a single twilight. Therefore, science flat-fields are typically generated by averaging twilight flats obtained over the course of the same observing run, which typically lasts no more than 10 days. This practice plausibly explains the similarity of spatial structure patterns across different fields.

Given the lack of clear documentation regarding the contributing observing runs and the involvement of another observatory and detector (T2KB) for a specific field, no further analysis is conducted based on the current combined photometry.

\subsection{The Result of Calibration}
\begin{figure*}[htbp]
\includegraphics[width=180mm]{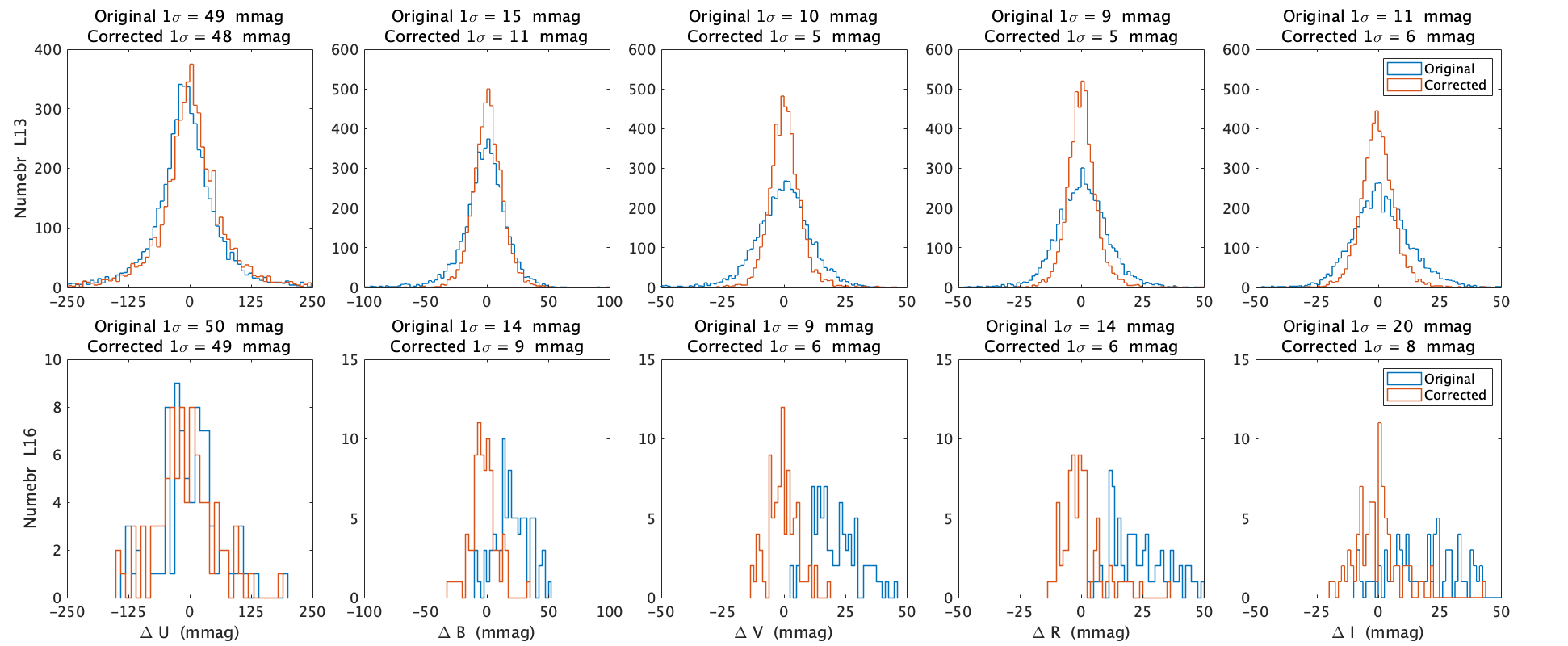}
\caption{The distributions of the magnitude differences between the original and corrected L13/L16 photometry and the standard stars ($G<16$). The standard deviations are indicated at the top of each panel. The first row presents the results for L13, and the second row presents those for L16.} 
\label{Final_precision}
\end{figure*}

After applying the corrections, we obtain the calibrated UBVRI magnitudes from the L13 and L16 datasets. 
Figure~\ref{Final_precision} shows the distributions of the magnitude differences between the original and corrected L13/L16 photometry and the standard stars.
Since the dispersion in each band is primarily dominated by the photometric uncertainties of both L13/L16 and the standard stars -- particularly for fainter sources, as demonstrated in the second and fourth rows of Figure~\ref{CMcor} -- only sources with $G < 16$ are included in Figure~\ref{Final_precision}.
The results show that our calibration significantly improves the consistency between L13/L16 and the standard stars, reducing the dispersion to approximately 11 mmag in the B band and to 5 -- 6 mmag in the VRI bands. In the U band, while no obvious improvement in dispersion is observed due to larger uncertainties of BEST stars, the distribution becomes more symmetric.

\subsection{Quality assessment for L92 standard stars}
We perform a quality assessment of the L92 by comparing them with the standardized BEST stars. A total of 268 common sources are found (positions linearly propagated to epoch J2000 using Gaia proper motions), among which 167 fall within the standardization ranges. 
According to \cite{Landolt1992a}, a diaphragm aperture of 14 arcsec in diameter was used for most observations, corresponding to a radius of 7 arcsec. To ensure that the Gaia measurements are consistent with those of L92, we visually inspect the PS1 DR1 (\citealt{PS1DR1}) images for common sources and confirm that no neighboring stars are present within this radius.

\begin{figure*}[htbp]
\includegraphics[width=180mm]{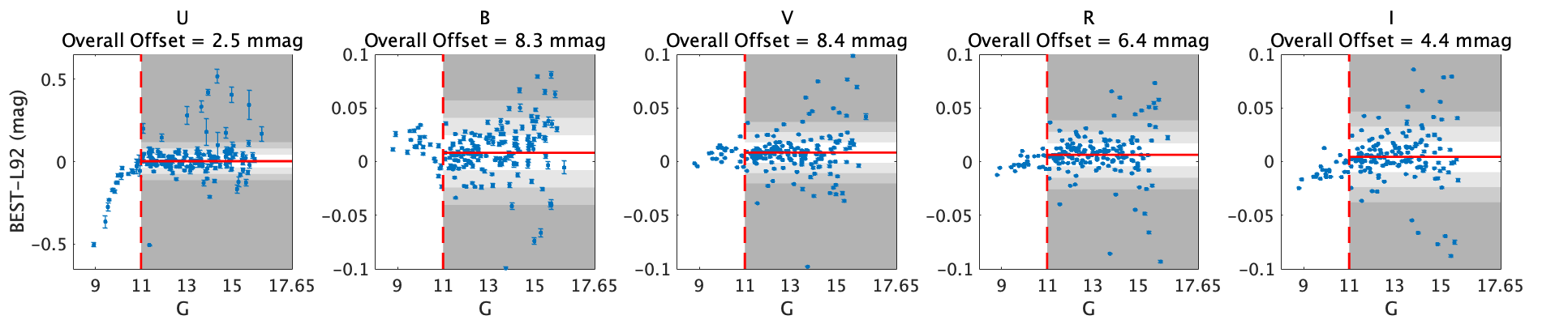}
\caption{The comparison between L92 and standardized BEST stars.
The error bars represent the uncertainties from BEST stars. The red vertical dashed line indicates the brightness limit of the standardization. The red horizontal solid line marks the overall offset. The shaded regions from white to gray indicate the ranges within 1$\sigma$ to 3$\sigma$ and beyond 3$\sigma$, respectively.} 
\label{L92_assessment}
\end{figure*}

The comparison results are shown in Figure~\ref{L92_assessment}.
As illustrated, the overall offsets in each band are smaller than 10 mmag, and the photometry of most common sources between L92 and the BEST stars is in good agreement. Even for those outside the standardization brightness range as indicated by the red vertical dashed line, the discrepancies are predominantly systematic, likely arising from extrapolation in the standardization procedure.
The shaded regions from white to gray indicate the ranges within 1$\sigma$ to 3$\sigma$, and beyond 3$\sigma$, respectively. Based on these comparisons, quality flags are assigned to each matched L92 source according to its consistency with the BEST standard stars, as described in Table~\ref{Table1}.

\begin{figure*}[htbp]
\includegraphics[width=180mm]{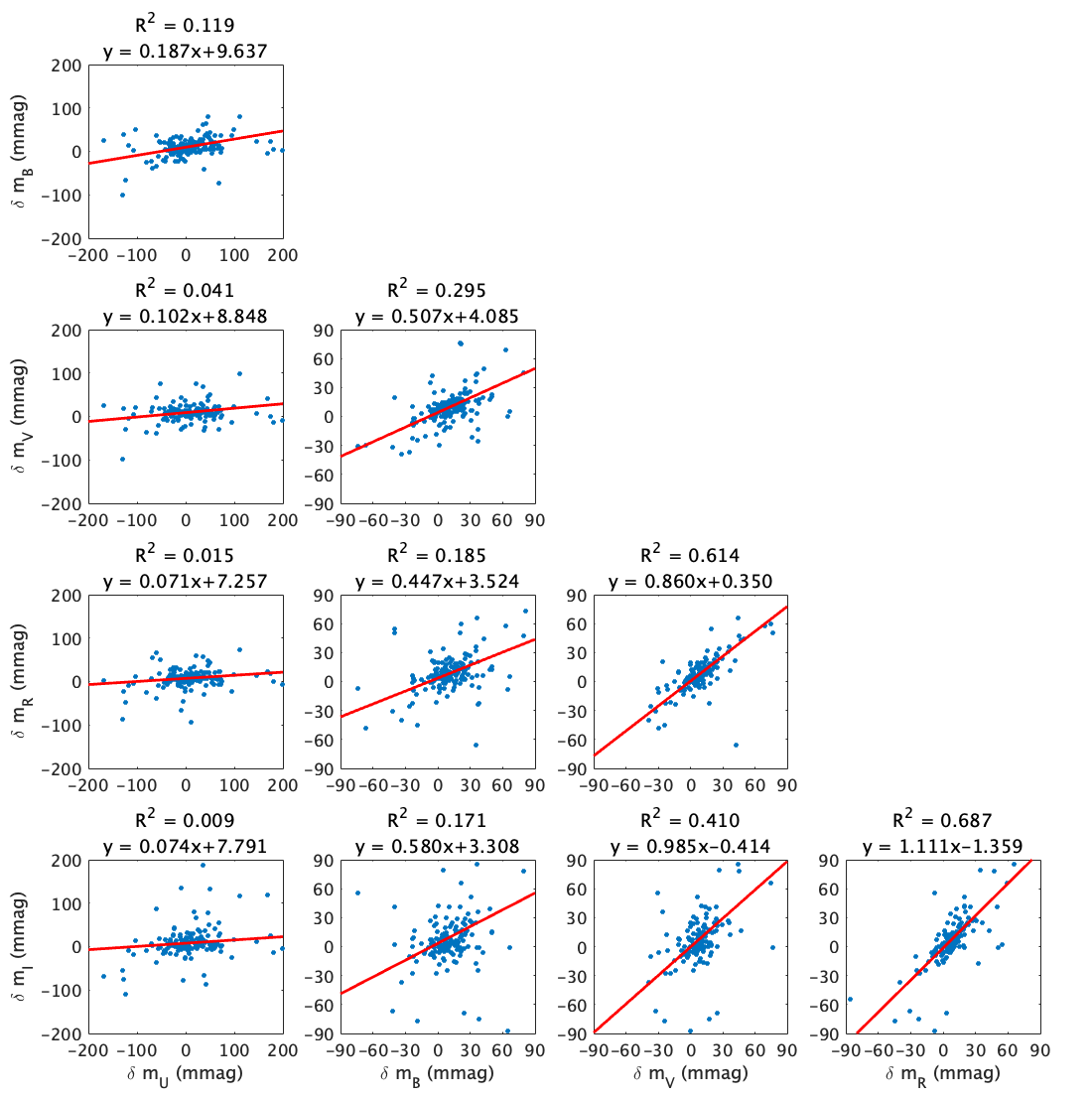}
\caption{The pairwise comparisons of the BEST$–$L92 differences across the UBVRI bands, presented with the same layout as Figure~\ref{FFZP_rela}.} 
\label{L92_rela}
\end{figure*}

We further examine the correlations between band pairs, as shown in Figure~\ref{L92_rela}. Similar to the zero-points of L13/L16, correlations are observed, particularly for pairs with closer effective wavelengths (e.g., $B$ vs.\ $V$, $V$ vs.\ $R$, $R$ vs.\ $I$, $V$ vs.\ $I$). The slopes of the linear regressions for these band pairs, , calculated using only sources with separations less than 7 arcsec, are comparable to those found in the L13/L16 zero-point comparisons, which may partly explain the zero-point correlations discussed in Section 3.3.

These correlations among L92 magnitudes are expected and are not a new finding of this work.
The correlations likely arise from the measurement procedures and subsequent transformations:
Firstly, as described by \cite{Landolt1992a}, each star is observed through a series of repeated measurements in the sequence VBURIIR (star plus sky), followed by VBURIIR (sky only). This observing strategy removes instrumental drift and variations in atmospheric transparency and sky brightness, suppresses scintillation and short-term variations, and ensures that residual atmospheric extinction coefficients are small and well determined. Consequently, colors can be defined with high precision. 
In these photomultiplier observations, only $V$ magnitude  is measured directly, and the other magnitudes are derived from this magnitude and the measured colors. This observational procedure inherently propagates any patterns from one band to the others.
Secondly, since both the filter system and the photomultiplier changed during the observations, a series of filter transformations were applied to data obtained at different epochs to ensure consistency. These transformations are typically carried out by first converting the V band, followed by colors such as $B-V$, $V-R$, and $R-I$. Consequently, any offset in a single band propagates into the adjacent bands. 
In summary, the systematic offsets observed across the $UBVRI$ bands for individual sources measured with a photomultiplier are likely attributable to both patterns present in the $V$ band and the color transformations. 
Here, we only identify and assess this issue; no further corrective procedures are applied.

We further examine the color residuals of the BEST$-$L92 common sources and find no significant error correlations among L92 colors, nor any other discernible systematic patterns in any of the colors. These findings further verify the internal consistency and robustness of the L92 colors.

Separately, an anonymous referee provided the following valuable assessment of the Landolt standard stars, which may be valuable for readers.
For stars in common between \cite{Landolt1983} and \cite{Landolt1992a}, the latter measurements may include observations from the former, implying that the two data sets are not statistically independent. Consequently, a simple weighted average would underestimate the uncertainties by roughly a factor of $\sqrt{2}$. For the use of standard stars, the \cite{Landolt1992a} results are generally preferred for stars in common, while the \cite{Landolt1983} data can be adopted for stars absent from the 1992 catalog.
Comparison with \cite{Landolt1973} shows that many UBV stars do not reappear in later catalogs, and the number of observations reported in 1973 does not correlate with those in 1992. On average, the 1992 measurements are brighter than the 1973 values by 0.0023 $\pm$ 0.0022 mag in U, 0.0016 $\pm$ 0.0012 mag in B, and 0.0017 $\pm$ 0.0014 mag in $V$ (weighted mean differences for 128 stars in common). Applying these offsets to the 1973 values and taking a weighted average with the 1992 results produces an enlarged set of UBV standards, with modestly improved precision for stars in common.

\subsection{Final catalog}
The calibrated catalog includes the original all photometry information from L13/L16/L92, Gaia phtotmetry and corrected UBVRI magnitudes (only for L13/L16) as described in Table~\ref{Table1}. 
As noted in Section 3.1, the final calibrated catalog may inherit certain issues, including variable stars, galaxies, and poor photometry from L13 and L16, which cannot be fully corrected by our calibration. Users should take these issues into account when using the catalog.

\renewcommand{\arraystretch}{0.7}
\setlength{\tabcolsep}{1.5mm}{
\begin{table}[htbp]
\footnotesize
\centering
\caption{Description of the Calibrated Catalog}
\begin{tabular}{lll}

\hline
\hline
Field & Description & Unit\\
\hline
Field & Field name & --\\
Star & Star number in the field & --\\
RAJ2000 & Right ascension in epoch J2000 & deg\\
DEJ2000 & Declination in epoch J2000 & deg\\
Umag & Original U magnitude & --\\
e\_Umag & Uncertainty in the U magnitude & mag\\
o\_Umag & Number of measurements in U & --\\
Bmag & Original B magnitude & --\\
e\_Bmag & Uncertainty in the B magnitude & mag\\
o\_Bmag & Number of measurements in B & --\\
Vmag & Original V magnitude & --\\
e\_Vmag & Uncertainty in the V magnitude & mag\\
o\_Vmag & Number of measurements in V & --\\
Rmag & Original R magnitude & --\\
e\_Rmag & Uncertainty in the R magnitude & mag\\
o\_Rmag & Number of measurements in R & --\\
Imag & Original I magnitude & --\\
e\_Imag & Uncertainty in the I magnitude & mag\\
o\_Imag & Number of measurements in I & --\\
U\_B & U$-$B color & --\\
e\_U\_B & Uncertainty in the U$-$B color & mag\\
B\_V & B$-$V color & --\\
e\_B\_V & Uncertainty in the B$-$V color & mag\\
V\_R & V$-$R color & --\\
e\_V\_R & Uncertainty in the V$-$R color & mag\\
R\_I & R\_I color & --\\
e\_R\_I & Uncertainty in the R$-$I color & mag\\
source\_id & Unique source identifier for Gaia EDR3 & --\\
ra & Right ascension from Gaia DR3 & deg\\
dec & Declination from Gaia DR3 & deg\\
parallax & parallax from Gaia EDR3 & mas\\
pmra & Proper motion in R.A. from Gaia EDR3 & mas/yr \\
pmdec &	Proper motion in Decl. from Gaia EDR3 & mas/yr \\
bp\_mag & Gaia EDR3 BP-band mean magnitude & -- \\
rp\_mag & Gaia EDR3 RP-band mean magnitude  & -- \\
g\_mag & Gaia EDR3 G-band mean magnitude & -- \\
ruwe & Gaia EDR3 Renormalized unit weight error & -- \\
excess\_factor & excess factor from Gaia EDR3 & -- \\
U\_corrected & Corrected U magnitude & --\\
B\_corrected & Corrected B magnitude & --\\
V\_corrected & Corrected V magnitude & --\\
R\_corrected & Corrected R magnitude & --\\
I\_corrected & Corrected I magnitude & --\\
is\_Calibrator$^{a}$ & Flag for calibrator stars;  & --\\
From$^{b}$ & Source literature of the star & --\\
Cali\_set$^{c}$ & Calibration form for fields & --\\
SS\_set$^{d}$ & Spatial structure correction form for stars & --\\
L92\_Assessment$^{e}$ & Quality assessment for L92 standard stars & --\\
\hline
\label{Table1}
\end{tabular}
\tablecomments{(a)
0 : Not a calibrator star;
1 : Is a calibrator star.}
\tablecomments{(b)
13 : Clem et al. 2013; 16 : for Clem et al. 2016; 92 : Landolt 1992.}
\tablecomments{(c)
0 : No correction applied;
1 : Only zero-point correction included;
2 : Both zero-point correction and spatial structure correction included, but with insufficient calibrator stars;
3 : Both zero-point correction and spatial structure correction included, with sufficient calibrator stars.}
\tablecomments{(d)
0 : No calibrator star for the spatial structure correction and no spatial structure correction applied;
1 : Only 1 to 5 calibrator stars for the spatial structure correction and no spatial structure correction applied;
2 : Only 6 to 9 calibrator stars for the spatial structure correction and spatial structure correction applied using the median;
3 : More than 10 calibrator stars for the spatial structure correction and spatial structure correction applied using the linear fitting.}
\tablecomments{(e)
0: Not assessed due to being unmatched or outside the standardization range;  
1: Located within the 1$\sigma$ range;  
2: Located within the 1 – 2$\sigma$ range;  
3: Located within the 2 – 3$\sigma$ range;  
4: Located beyond the 3$\sigma$ range.
The five digits indicate the U, B, V, R, and I bands respectively.}

\end{table}}
\renewcommand{\arraystretch}{1}

\section{The Generation of LAMOST based UBVRI Standard stars}
As described in \cite{S82Yuan2015a} and \cite{Huang2022aS82}, the SCR method enables precise determination of intrinsic colors from a few physical parameters. Consequently, the stellar atmospheric parameters provided by LAMOST DR12, in conjunction with the Gaia DR3 BP/RP photometry as a precise indicator of stellar brightness, can be employed to derive photometry in the Johnson and Kron-Cousins UBVRI bands using the following equations:
\begin{equation}
    XP-X = (XP-X)_0 + E(BP-RP) \times R(XP-X)
    \label{EQ2}
\end{equation}
\begin{equation}
    (XP-X)_0 = f(T_{eff},log\,g,[Fe/H])
    \label{EQ3}
\end{equation}
Here $XP - X$ denotes the color combinations $BP - U$, $BP - B$, $RP - V$, $RP - R$, and $RP - I$, respectively. The term $R(XP - X)$ refers to the extinction coefficient for each color relative to extinction $E(BP - RP)$. In this notation, $XP$ represents either $BP$ or $RP$, and $X$ refers to one of the Johnson or Kron-Cousins $UBVRI$ bands, with the $BP$ and $RP$ photometry incorporating the magnitude-dependent corrections described in \cite{YL2021}.
According to the above equations, once the precise $E(BP - RP)$ extinction for each source, the common extinction coefficients $R(XP - X)$, and the transformation relation $f$ are determined, the UBVRI magnitudes for each source can be derived accordingly.

\subsection{Precise $E(BP-RP)$ Extinction and Temperature- and Extinction-dependent UBVRI Extinction Coefficients}

Precise extinction correction is essential for the SCR method and the first requirement for achieving it is obtaining an precise extinction value. We avoid using any dust-reddening map, as they are either insufficiently precise or, like SFD98 (\citealt{SFD98}), unreliable at low Galactic latitudes and affected by spatially dependent systematic errors (\citealt{SunSFD}). Instead, we adopt star-pair technique (\citealt{starpairYuan2013} and \citealt{starpairZhang2020}) to determine the values of $E(BP - RP)$, utilizing LAMOST stellar atmospheric parameters and precise Gaia $BP-RP$ color.
Similarly, the values of $E(XP - X)$ and intrinsic color $(XP-X)_0$ for each color combination can be determined using the star-pair technique, by replacing the color with the corresponding standardized BEST XPSP color.
Given the limited number of common sources between the LAMOST and XPSP samples, and the widespread availability of BP/RP photometry for LAMOST stars, extinction coefficients $R(XP - X)$ are necessary to convert $E(BP - RP)$ into $E(XP - X)$ for individual LAMOST sources.

To maintain a uniform distribution in stellar parameters and ensure the reliability of the sample, we apply both the standardization limits and the parameter constraints $4250 \mathrm{K} \le T_{\mathrm{eff}} \le 8000 \mathrm{K}$, $0.5 \le \log g \le 4.9$, and $-2.5 \le \mathrm{[Fe/H]} \le 0.5$ to the final set of LAMOST \& Gaia-based standard stars.
Due to the color limitations of the BEST XPSP standardization, cool stars with high extinction are excluded. Although this exclusion does not affect the present results, it reduces the general applicability of the derived extinction coefficients. To mitigate this, we supplement the sample with cool common sources between LAMOST and the calibrated L13/L16 data.

\begin{figure*}[htbp]
\includegraphics[width=180mm]{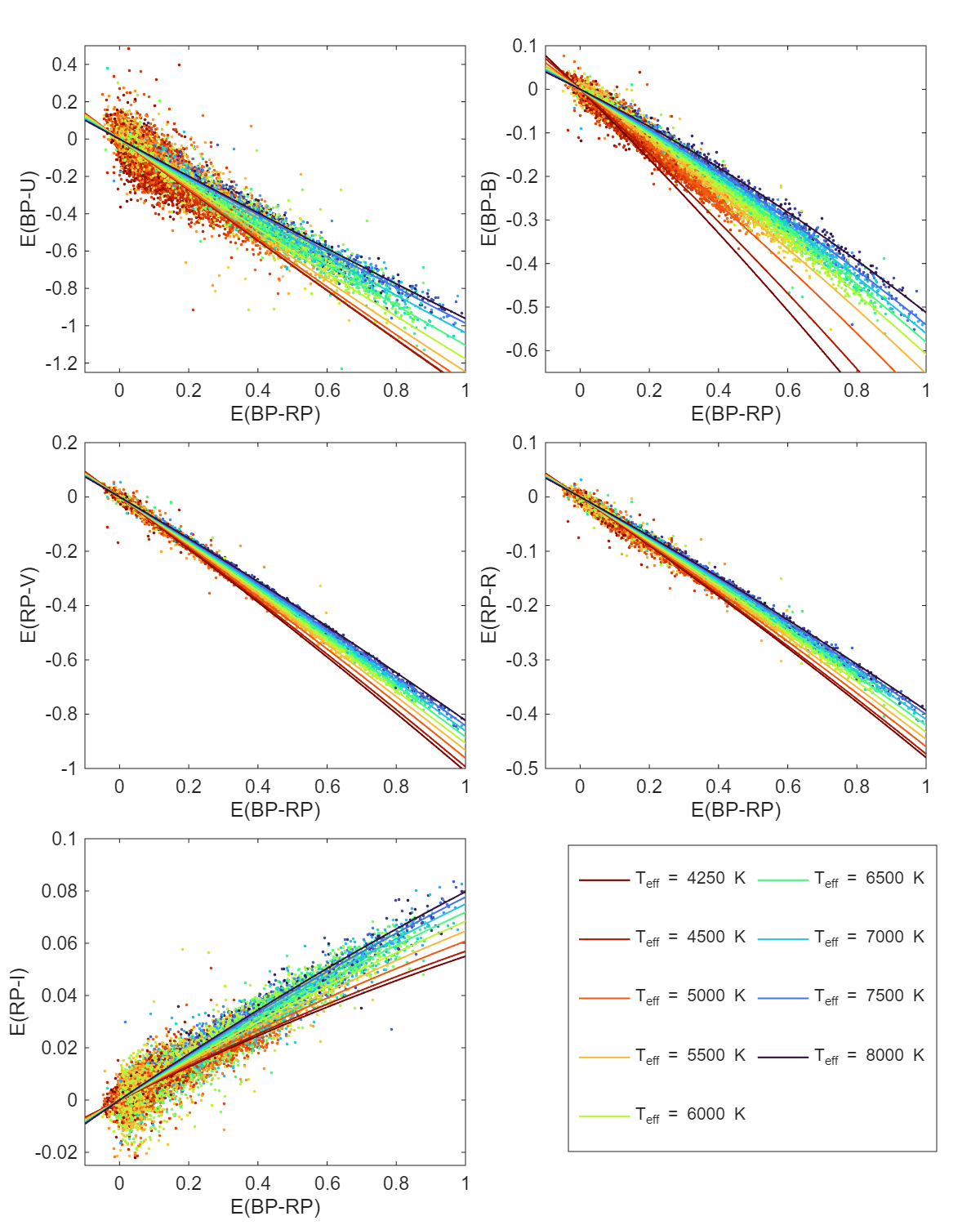}
\caption{The fitting of temperature- and extinction-dependent UBVRI extinction coefficients for $E(XP - X)$. The solid curves, ranging from red to blue, represent the fitted results for temperatures from 4250 K to 8000 K, respectively. The color coding of the data points matches that of the curves.} 
\label{Extinction_Coefficient}
\end{figure*}

The resulting extinction coefficients are presented in Figure~\ref{Extinction_Coefficient}, with the corresponding numerical values listed in Table~\ref{Table2}. For reference, we also provide single-valued extinction coefficients for $T_{\mathrm{eff}} = 5000 \mathrm{K}$, $T_{\mathrm{eff}} = 5500 \mathrm{K}$ and $T_{\mathrm{eff}} = 6000 \mathrm{K}$, valid for cases with $E(BP-RP) \le 0.5$. A machine-readable version of Table~\ref{Table2} will also be available at \url{https://doi.org/10.12149/101704}.

It should be emphasized that only the coefficients for converting $E(BP - RP)$ to $E(XP - X)$ are derived directly in this work. The single-band coefficients for BP, RP, and UBVRI are obtained from, or combined with, the results of \cite{ZRY2023a}, which are valid for $E(B - V) \le 0.5$.
For applications requiring extinction coefficients referenced to $E(B-V)$, we recommend first deriving $E(BP - RP)$ using the coefficients provided in \cite{ZRY2023a}, since the extinction coefficients depend on the extinction values themselves, rendering it challenging to express the $E(B-V)$-referenced coefficients in a simple explicit form. 
It should be noted that the current extinction coefficients represent only the average results within the LAMOST region. Variations in properties of interstellar dust, along different sight lines can lead to differences in the interstellar extinction law, and consequently affect the extinction coefficients. For a detailed discussion, see Section 4.3.

\setlength{\tabcolsep}{1.5mm}{
\begin{table*}[htbp]
\footnotesize
\centering
\caption{Single-valued Extinction Coefficients and Fitting Coefficients of $R_{E(BP-RP)}(T_{\mathrm{eff}} ,(E(BP-RP)))$}
\begin{tabular}{c c c c c c c c c c}

\hline
\hline
Color & $R_{5000 \mathrm{K}}$ & $R_{5500 \mathrm{K}}$ & $R_{6000 \mathrm{K}}$ & $C_1$ & $C_2$ & $C_3$ & $C_4$ & $C_5$ & $C_6$\\
\hline

$BP-U$ & $-$1.328 & $-$1.273 & $-$1.205 & $-$1.330 $\times 10^{-11}$ & \,\,\,\,2.475 $\times 10^{-7}$ & $-$0.001388 & 0.0000 & \,\,\,\,0.04284 & \,\,\,\,1.071 \\
$BP-B$ & $-$0.6555 & $-$0.5865 & $-$0.5421 & \,\,\,\,1.095 $\times 10^{-11}$ & $-$2.299 $\times 10^{-7}$ & \,\,\,\,0.001646 & 0.0000 & $-$0.1060 & $-$4.467 \\
$RP-V$ & $-$0.9175 & $-$0.8888 & $-$0.8629 & \,\,\,\,5.848 $\times 10^{-13}$ & $-$1.530 $\times 10^{-8}$ & \,\,\,\,0.0001697 & 0.0000 & $-$0.07102 & $-$1.430 \\
$RP-R$ & $-$0.4344 & $-$0.4208 & $-$0.4077 & $-$5.682 $\times 10^{-13}$ & \,\,\,\,8.523 $\times 10^{-9}$ & $-$1.534 $\times 10^{-5}$ & 0.0000 & $-$0.04052 & $-$0.4845 \\
$RP-I$ & \,\,\,\,0.06737 & \,\,\,\,0.07124 & \,\,\,\,0.07498 & $-$1.618 $\times 10^{-13}$ & \,\,\,\,2.427 $\times 10^{-9}$ & $-$4.368 $\times 10^{-6}$ & 0.0000 & $-$0.01056 & \,\,\,\,0.05273 \\
$BP^*$ & \,\,\,\,2.472 & \,\,\,\,2.424 & \,\,\,\,2.384 & $-$6.534 $\times 10^{-13}$ & \,\,\,\,2.694 $\times 10^{-8}$ & $-$0.0003253 & $-$0.32 & \,\,\,\,0.4226 & \,\,\,\,3.396 \\
$RP^*$ & \,\,\,\,1.472 & \,\,\,\,1.424 & \,\,\,\,1.384 & $-$6.534 $\times 10^{-13}$ & \,\,\,\,2.694 $\times 10^{-8}$ & $-$0.0003253 & $-$0.32 & \,\,\,\,0.4226 & \,\,\,\,2.396 \\
$U^*$ & \,\,\,\,3.800 & \,\,\,\,3.752 & \,\,\,\,3.588 & \,\,\,\,1.265 $\times 10^{-11}$ & $-$2.206 $\times 10^{-7}$ & \,\,\,\,0.001063 & $-$0.32 & \,\,\,\,0.3797 & \,\,\,\,2.325 \\
$B^*$ & \,\,\,\,3.128 & \,\,\,\,3.079 & \,\,\,\,2.926 & $-$1.160 $\times 10^{-11}$ & \,\,\,\,2.569 $\times 10^{-7}$ & $-$0.001971 & $-$0.32 & \,\,\,\,0.5286 & \,\,\,\,7.863 \\
$V^*$ & \,\,\,\,2.389 & \,\,\,\,2.341 & \,\,\,\,2.246 & $-$1.238 $\times 10^{-12}$ & \,\,\,\,4.224 $\times 10^{-8}$ & $-$0.0004949 & $-$0.32 & \,\,\,\,0.4936 & \,\,\,\,3.826 \\
$R^*$ & \,\,\,\,1.906 & \,\,\,\,1.858 & \,\,\,\,1.791 & $-$8.524 $\times 10^{-14}$ & \,\,\,\,1.842 $\times 10^{-8}$ & $-$0.0003099 & $-$0.32 & \,\,\,\,0.4631 & \,\,\,\,2.881 \\
$I^*$ & \,\,\,\,1.405 & \,\,\,\,1.356 & \,\,\,\,1.309 & $-$4.916 $\times 10^{-13}$ & \,\,\,\,2.451 $\times 10^{-8}$ & $-$0.0003209 & $-$0.32 & \,\,\,\,0.4331 & \,\,\,\,2.343 \\

\hline

\label{Table2}
\end{tabular}
\tablecomments{The function forms are 
$R_{E(BP-RP)}(T_{\mathrm{eff}} ,(E(BP-RP)))\,=\, C_1 \times T_{\mathrm{eff}} ^3 + C_2 \times T_{\mathrm{eff}} ^2 + C_3 \times T_{\mathrm{eff}}+C_4 \times E(BP-RP) ^2 +C_5 \times E(BP-RP) +C_6 $.}
\tablecomments{($^*$) Coefficients derived from or combined with the results of \cite{ZRY2023a}.}
\end{table*}}

\subsection{Result and Comparison}
\begin{figure}[htbp]
\includegraphics[width=90mm]{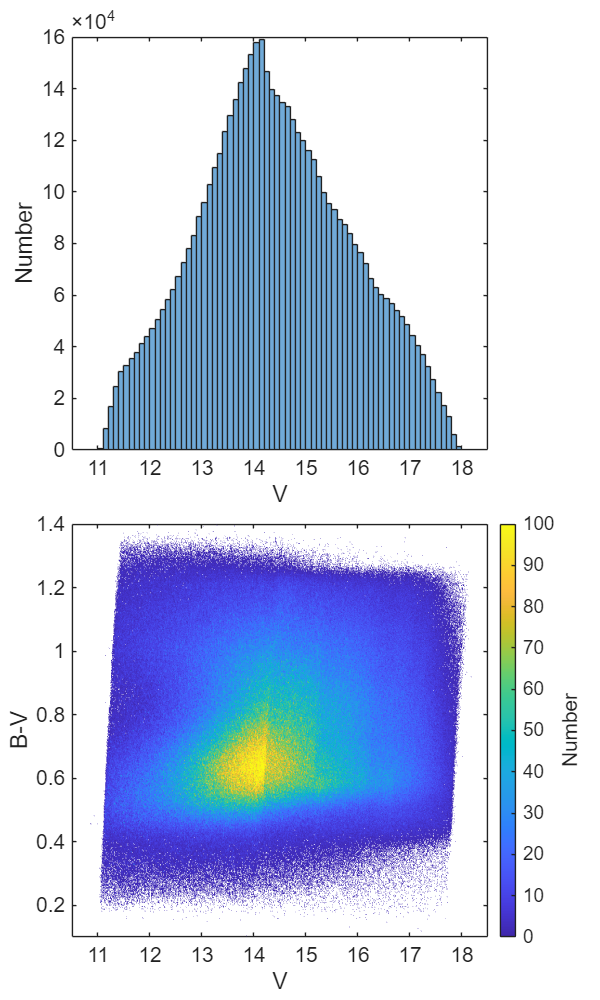}
\caption{The brightness and color distribution of final LAMOST \& Gaia-based standard star catalog, color-coded by density.} 
\label{LMdistribution}
\end{figure}

The final LAMOST \& Gaia-based standard star catalog contains approximately 5.4 million sources with signal-to-noise ratios in the g band of the LAMOST spectra greater than 15. As shown in the Figure~\ref{LMdistribution}, the standard stars are predominantly distributed within a brightness range of $V \sim $  11 -- 18 and a color range of $B-V \sim $ 0.2 -- 1.4.

\begin{figure*}[htbp]
\includegraphics[width=180mm]{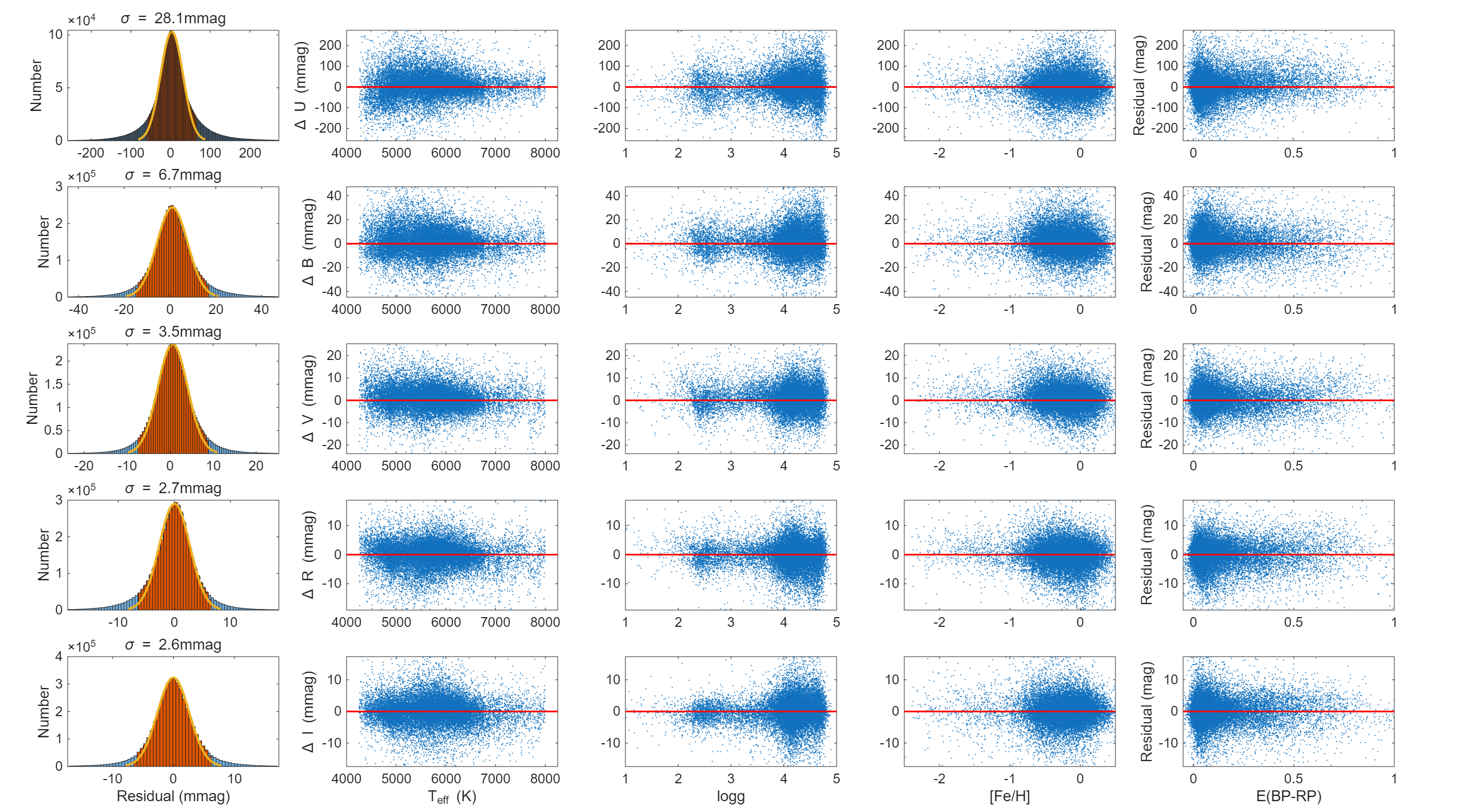}
\caption{The comparison between LAMOST \& Gaia-based UBVRI photometry and the corresponding BEST XPSP standard stars in the UBVRI bands. The leftmost column illustrates the distributions of the differences, where the red bar marks the range used to compute the standard deviation of the Gaussian fit, focusing on the central distribution. The second to fifth columns present the differences as functions of $T_{\mathrm{eff}}$, log g, [Fe/H], and $E(BP-RP)$, respectively.} 
\label{BESTLM_compare}
\end{figure*}

\begin{figure*}[htbp]
\includegraphics[width=180mm]{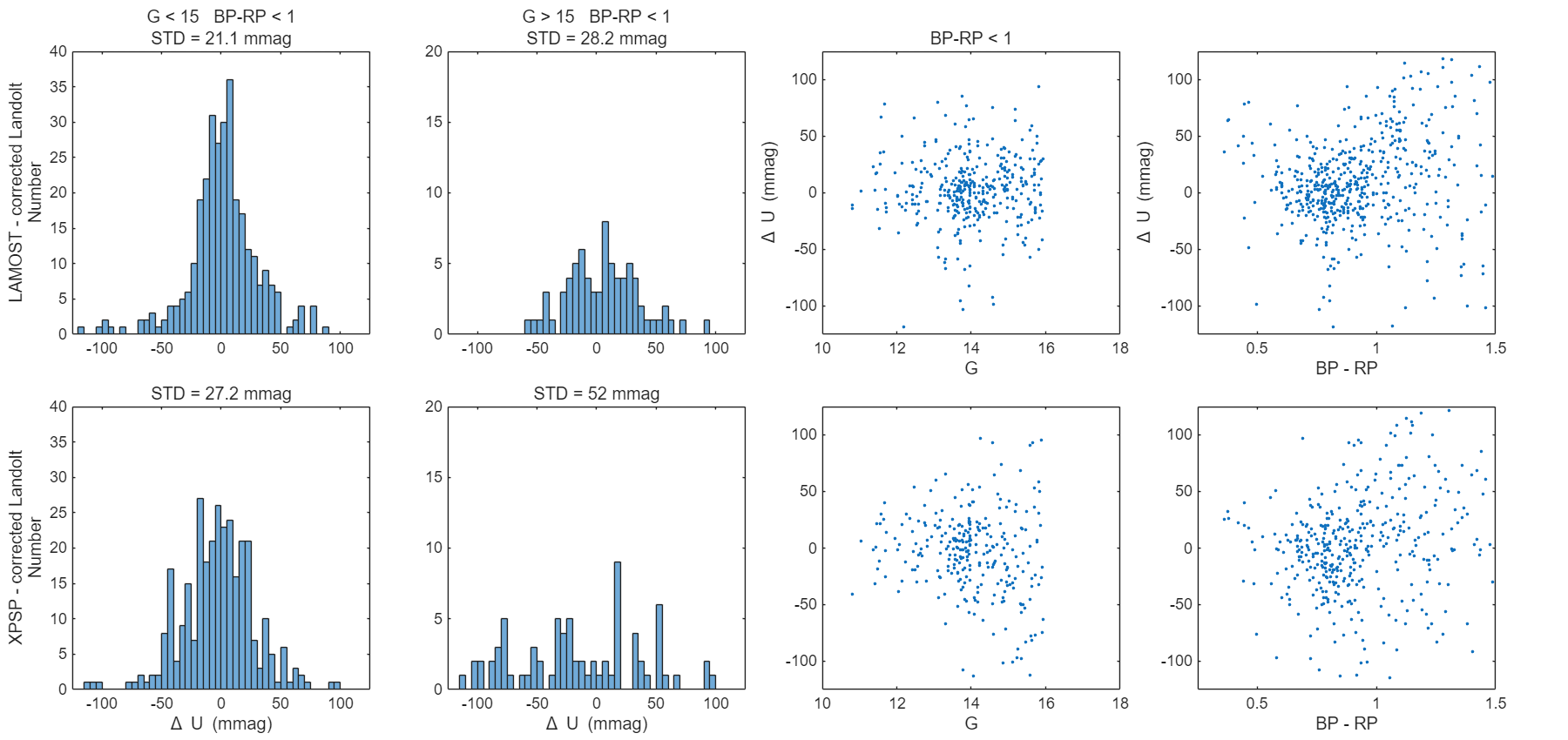}
\caption{The comparison of U band photometry between LAMOST \& Gaia-based catalog (top panels) and BEST XPSP catalog (bottom panels) against the Landolt standard stars. The left two columns show histograms of differences for brighter and fainter stars with color $BP–RP < 1$, with the corresponding standard deviations (STD) indicated in each case. The right two panels present the differences as a function of $G$ and color $BP - RP$.} 
\label{crosscheck}
\end{figure*}

The final catalog provides Gaia photometry, LAMOST \& Gaia-based UBVRI magnitudes, and extinction values, as summarized in Table~\ref{Table3}.
To evaluate the precision, star-pair results, and extinction correction of our LAMOST \& Gaia-based UBVRI photometry, we compare it with the corresponding BEST XPSP standard stars. As shown in Figure~\ref{BESTLM_compare}, the agreement reaches 28 mmag in the $U$ band, 7 mmag in the $B$ band, and about 3 mmag in the $VRI$ bands, primarily reflecting the precision for dwarfs with effective temperatures between approximately 4500 K and 6500 K. No significant systematic trends are observed as functions of the LAMOST stellar atmospheric parameters or extinction, indicating that both our star-pair results and extinction correction are reliable. 
We further use the corrected Landolt standard stars to independently assess the LAMOST \& Gaia-based and XPSP U-band photometry. As shown in Figure~\ref{crosscheck}, for sources with $BP-RP < 1$ -- which includes the vast majority of stars -- the LAMOST \& Gaia-based U-band photometry achieves noticeably higher precision than XPSP, and this precision advantage becomes even more pronounced for fainter stars.

\setlength{\tabcolsep}{1.5mm}{
\begin{table}[htbp]
\footnotesize
\centering
\caption{Description of the LAMOST-based standard star catalog}
\begin{tabular}{lll}

\hline
\hline
Field & Description & Unit\\
\hline
source\_id & Unique source identifier for Gaia EDR3 & --\\
ra & Right ascension from Gaia DR3 & deg\\
dec & Declination from Gaia DR3 & deg\\
parallax & parallax from Gaia EDR3 & mas\\
pmra & Proper motion in R.A. from Gaia EDR3 & mas/yr \\
pmdec &	Proper motion in Decl. from Gaia EDR3 & mas/yr \\
bp\_mag & Gaia EDR3 BP-band mean magnitude & -- \\
rp\_mag & Gaia EDR3 RP-band mean magnitude  & -- \\
g\_mag & Gaia EDR3 G-band mean magnitude & -- \\
ruwe & Gaia EDR3 Renormalized unit weight error & -- \\
excess\_factor & excess factor from Gaia EDR3 & -- \\
U\_LAMOST & LAMOST-based U magnitude & --\\
B\_LAMOST & LAMOST-based B magnitude & --\\
V\_LAMOST & LAMOST-based V magnitude & --\\
R\_LAMOST & LAMOST-based R magnitude & --\\
I\_LAMOST & LAMOST-based I magnitude & --\\
$E(B-V)$ & Extinction for $B-V$ color from SFD98 & --\\
$E(BP-RP)$ & Extinction for $BP-RP$ color from this work & --\\
\hline
\label{Table3}
\end{tabular}
\end{table}}

\subsection{Limitations}
\begin{figure*}[htbp]
\includegraphics[width=180mm]{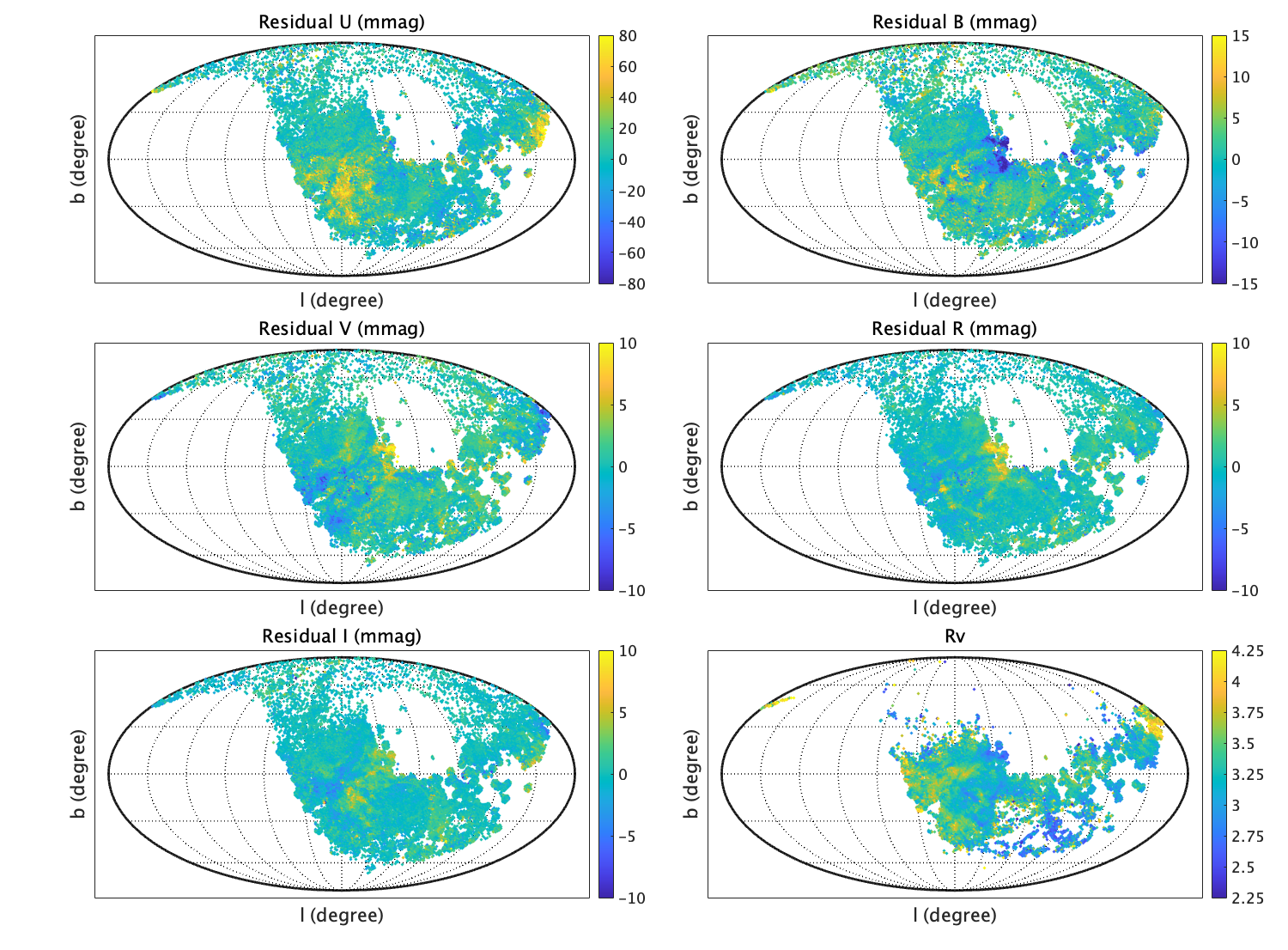}
\caption{The spatial distribution of the differences between the LAMOST \& Gaia-based UBVRI photometry and the corresponding BEST XPSP values in the UBVRI bands. The panel at the bottom right shows the distribution of $Rv$, taken from \cite{ZRY2023b}. Each panel presents the mean values, with the celestial sphere partitioned into 196,608 regions using the HEALPix scheme (nside = 128, corresponding to a spatial resolution of approximately 27.5′).}
\label{Figure_Rv}
\end{figure*}

The LAMOST \& Gaia-based UBVRI standard star catalog is primarily designed for practical photometric calibration rather than detailed studies of individual stars, as the physical information contained in the derived UBVRI magnitudes is already explicitly represented by the stellar atmospheric parameters from LAMOST and the BP/RP photometry from Gaia. Accordingly, the catalog is constructed using a simplified stellar model $(XP-X)_0 = f(T_{eff},log\,g,[Fe/H])$ and assumes that the extinction coefficients are spatially constant across all bands.
The results in Section 4.2 show good overall consistency between the current catalog and the BEST XPSP standard stars, indicating that the simplified assumptions are statistically valid for large samples. However, several effects may cause deviations of the actual UBVRI magnitudes from the simplified $(XP-X)_0 = f(T_{eff},log\,g,[Fe/H])$ model, which likely manifest as non-Gaussian tails in the comparison with BEST XPSP.
These potential effects include:
\begin{enumerate}
    \item Unresolved binaries: The current catalog is constructed under the single-star assumption. The photometric results of most unresolved binaries based on the SCR method are expected to show only minor deviations. And binaries with specific mass ratios between the two components can contribute to the non-Gaussian long tails when compared with BEST. For unresolved binaries, the observable spectrum of a system is the superposition of two single-star spectra. Stellar atmospheric parameters derived from single-star models can therefore exhibit modest systematic biases (\citealt{BinarySpec}). The effects of metallicity and log $g$ on the intrinsic color generation of main-sequence binaries are minor, and both the $T_{eff}$ values from spectra and the color from photometry generally fall between those of the primary and secondary stars. According to \cite{Yuan2015c}, the colors of binaries deviate slightly from the stellar loci of single stars; only under specific color combinations of the two components can the maximum offsets reach approximately 0.15 mag, 0.036 mag, $-$0.036 mag, and $-$0.036 mag in the SDSS $u-g$, $g-r$, $r-i$, and $i-z$ colors, respectively. The deviations are nearly zero when the mass ratios of the two components are close to unity or highly unequal.
    \item Rotational effects: The colors of rapidly rotating stars may vary depending on whether they are viewed equator-on or pole-on. Since the current LAMOST sample includes only FGK-type stars, whose rotation speeds are generally low, this effect is expected to be minor.
    \item Molecular-band variations: Late-type stars with identical temperature, metallicity, and gravity can exhibit significant differences in molecular-band strengths. Such effects are minimized in our catalog, as low-temperature M-type stars have been excluded.
    \item Spatial variation of interstellar extinction law: The current extinction coefficients represent only the mean values over the LAMOST region, ignoring variations in the interstellar extinction law due to different dust properties along distinct lines of sight. Figure\,\ref{Figure_Rv} shows the spatial distribution of the differences between our catalog and BEST XPSP. Spatial structures are clearly visible in low Galactic latitude regions across all bands, which are associated with variations in $R_V$ (\citealt{ZRY2023b}). The corresponding amplitudes are approximately 11, 2.6, 1.5, 1.1, and 1.0 mmag in the U, B, V, R, and I bands, respectively, with maximum values reaching about 60, 10, 5, 5 and 5 mmag in these bands. In addition, arc-shaped patterns, especially visible in the $B$ band, are clearly associated with Gaia's scanning law and likely originate from differences between Gaia photometry and spectra. This spatial patterns are more likely attributed to the Gaia DR3 XP spectra, particularly in the $BP$ band (\citealt{HBW2024b}). These results suggest that variations in the extinction law do not significantly affect the statistical results, though local deviations may exist in specific sky regions.
\end{enumerate}

\section{Summary}
We perform a comprehensive recalibration of the Landolt standard stars in the Johnson $UBV$ and Kron–Cousins $RI$ systems for L13/L16, revealing zero-point offsets and spatial structures and significantly improving precision, while also assessing the quality of the L92 standard stars. The main findings of this work are summarized below.

Firstly, we identify an overall zero-point offset between L13 and L16, along with field-specific shifts of 5--14 mmag across $UBVRI$ bands, with correlated offsets between $BVRI$ bands. This correlation is likely due to the L92 based calibration process and the combined result of temporal zero-point drifts and the photometric transformation process. 

Secondly, spatial structures up to 7 -- 10 mmag are also detected in the $BVRI$ bands of most of fields with sufficient standard stars in L13 and L16, showing consistent patterns across bands within the same field and across fields within the same band, likely caused by the application of averaged flat fields per observing run. Such characterization of spatial structures up to 20 -- 30 mmag in $U$ band are performed only for two fields -- LSE259 and WD0830-535, which include sufficiently abundant standard stars, due to the high uncertainty of standard stars.

Thirdly, the examination shows that our calibration significantly improves the consistency between L13/L16 and the standard stars, reducing the dispersion to approximately 10 mmag in the B band and to 5 -- 6 mmag in the VRI bands for sources with $G < 16$. 

Finally, a further quality assessment for L92 standard stars of L92 reveals correlated discrepancies for individual stars between L92 and standard star dataset across the BVRI bands, which exhibit a similar slope with field-specific zero-point shifts correlation in L13 and L16 especially for $B$ vs.\ $V$, $V$ vs.\ $R$, $R$ vs.\ $I$. 
The correlations among L92 magnitudes are expected and do not constitute a new result of this study.
The correlated discrepancies for individual stars likely arise from the photomultiplier multi-band measurement procedures, which provide high-precision color measurements but measure only one magnitude directly, with the remaining magnitudes derived from this magnitude and the measured colors. This observational procedure inherently propagates any patterns from one band to the others. 
In addition, the discrepancies may also be contributed by the color transformations between two sets of filter systems and the photomultipliers.
While correlated discrepancies are present among the magnitudes, the colors of L92 show no significant error correlations or other systematic trends, confirming the internal consistency and reliability of the L92 colors.

Furthermore, from LAMOST DR12 stellar atmospheric parameters, Gaia DR3 photometry, and XPSP data, we derive temperature- and extinction-dependent extinction and reddening coefficients for $UBVRI$ bands, and construct a LAMOST \& Gaia-based catalog of 5.4 million $UBVRI$ new standard stars. This catalog contains an additional 0.6 million sources absent from XPSP, and for the vast majority of stars, the LAMOST \& Gaia-based U-band photometry attains superior precision relative to XPSP. The recalibrated Landolt standards and the LAMOST \& Gaia-based catalog will be released via the BEST website (\url{https://nadc.china-vo.org/data/best/}) and (\url{https://doi.org/10.12149/101704}).

\vspace{7mm} \noindent {\bf Acknowledgments}
The authors thank the anonymous referee for their suggestions, which improved the quality of this manuscript and offered important insights into the understanding of the Landolt/Clem standard stars.
This work is supported by the National Natural Science Foundation of China through the projects NSFC 12222301, 12173007 and 124B2055, and by the National Key R\&D Program of China under grants 2024YFA1611901 and 2024YFA1611601.
K.X. acknowledges support for the National Natural Science Foundation of China grants No. 12403024; the Postdoctoral Fellowship Program of CPSF under Grant Number GZB20240731; the Young Data Scientist Project of the National Astronomical Data Center; and the China Postdoctoral Science Foundation No. 2023M743447.

This work has made use of data from the European Space Agency (ESA) mission {\it Gaia} (\url{https://www.cosmos.esa.int/gaia}), processed by the Gaia Data Processing and Analysis Consortium (DPAC, \url{https:// www.cosmos.esa.int/web/gaia/dpac/ consortium}). Funding for the DPAC has been provided by national institutions, in particular the institutions participating in the Gaia Multilateral Agreement. 

The Guoshoujing Telescope (the Large Sky Area Multi-Object Fiber Spectroscopic Telescope LAMOST) is a National Major Scientific Project built by the Chinese Academy of Sciences. Funding for the project has been provided by the National Development and Reform Commission. LAMOST is operated and managed by the National Astronomical Observatories, Chinese Academy of Sciences.

{}

\end{CJK*}
\end{document}